\documentclass[preprint,12pt]{elsarticle}

\usepackage{amssymb}
\usepackage{amsmath}
\usepackage{amsthm}

\usepackage{booktabs}
\usepackage{tabularx}
\usepackage{array}
\usepackage{comment}
\usepackage{placeins}

\newcolumntype{Y}{>{\raggedright\arraybackslash}X}

\theoremstyle{plain}

\newtheorem{proposition}{Proposition}

\newtheorem{corollary}{Corollary}

\theoremstyle{definition}
\newtheorem{definition}{Definition}

\theoremstyle{remark}

\begin{document}

\begin{frontmatter}



\title{Generalized FDNF fuzzification of elementary cellular automata and its nonlinear pattern dynamics}


\author{Takiko Sasaki\fnref{fn1}} %% Author name

\affiliation[fn1]{organization={Department of Mathematical Engineering, Faculty of Engineering, Musashino University},%Department and Organization
            addressline={3-3-3 Ariake, Koto-ku}, 
            city={Tokyo},
            postcode={135-8181}, 
            country={Japan}}

\author{Seiryu Shimizu\fnref{fn1}} %% Author name

\author{Tetsuji Tokihiro\corref{cor1}%
\fnref{fn1}} 
\ead{t-toki@musashino-u.ac.jp}
\cortext[cor1]{Corresponding author}

%% Abstract
\begin{abstract}
Fuzzy disjunctive normal form (FDNF) gives the canonical multi-affine extension of an elementary cellular automaton (ECA) rule to the unit cube. Although it preserves the Boolean rule on binary states, its multi-affine structure can smooth high-contrast CA patterns and restrict continuous-state dynamics.

We introduce generalized FDNF rules
\[
\widetilde f_k^{g,u,v,w}(x,y,z)
=
g\left(f_k(u(x),v(y),w(z))\right),
\]
where the transformations \(g,u,v,w:[0,1]\to[0,1]\) fix the endpoints. The identity maps recover ordinary FDNF, while threshold-like, discontinuous, non-monotone, and expanding choices yield rule-preserving fuzzy ECAs.

We demonstrate, in representative rules, that the transformation shape strongly affects pattern dynamics: threshold-like maps promote ECA-like pattern recovery, parameter deformations interpolate toward FDNF-like smoothing, and discontinuities induce gap-generated regimes. Pattern changes are summarized by contrast, fuzziness, and a finite-resolution participation-type support exponent. In three-cell systems, an expanding non-monotone transformation yields stable period-six cycles for rule 210, verified by interval arithmetic, coexisting with an expanding invariant line set; rules 51 and 85 inherit one-dimensional expanding dynamics. The framework provides a rule-preserving bridge from Boolean cellular automata to fuzzy and continuous-state nonlinear dynamics.
\end{abstract}


\begin{comment}
\begin{highlights}
\item FDNF is formulated as the canonical multi-affine ECA extension.
\item Endpoint-preserving transformations yield generalized FDNF fuzzy rules.
\item Endpoint-preserving transformations affect pattern recovery and smoothing.
\item A participation-type support exponent quantifies finite-resolution amplitude-support changes.
\item Three-cell systems exhibit stable cycles, invariant line-set dynamics, and finite-resolution class structures.
\end{highlights}
\end{comment}

%% Keywords

\end{frontmatter}



% section 1
%%%%%%%%%%%%%%%%%%%%%%%%%%%%%%%%%%%
\section{Introduction}
\label{sec1}
%% Labels are used to cross-reference an item using \ref command.

Cellular automata (CAs) are discrete-time dynamical systems consisting of locally interacting cells, each of which takes values in a finite set \cite{vonNeumann1966}. Although their local transition rules are often very simple, CAs can generate highly nontrivial global behavior, including propagating structures, self-similar patterns, and irregular spatiotemporal dynamics \cite{Wolfram2002}.

An elementary cellular automaton (ECA) is a one-dimensional, two-state, radius-one cellular automaton \cite{Wolfram1986}. Each site has state 0 (off/white) or 1 (on/black), and its next state is determined by the states of the left, center, and right sites. 
ECAs have been used not only as basic models of pattern formation but also as idealized models for transport phenomena, including traffic-flow dynamics \cite{Nagel1992}.

A limitation of Boolean CAs is that each cell state is restricted to a finite set, typically \(\{0,1\}\). In many modeling situations, however, cell states may represent uncertain, ambiguous, or continuously varying quantities. 
Fuzzy cellular automata (FCAs) address this issue by extending the state space of a CA from a discrete set to a continuous interval, usually \([0,1]\) \cite{cattaneo1997}. 
For Boolean CAs, one of the most systematic algebraic approaches to fuzzification is based on fuzzy disjunctive normal form (FDNF) \cite{Betel2009}. 
In this construction, a Boolean local rule is first written in disjunctive normal form and then Boolean operations are replaced by fuzzy operations. For an ECA rule
\[
F_k:\{0,1\}^3\to\{0,1\},
\]
the resulting FDNF rule is a polynomial map
\[
f_k:[0,1]^3\to[0,1]
\]
that agrees with \(F_k\) on the eight vertices of the unit cube.

From a mathematical viewpoint, FDNF has a distinguished status: it is the unique multi-affine extension of the Boolean local rule to the unit cube.
Thus, FDNF is not merely an ad hoc fuzzification; it is a canonical interpolation of the vertex values of a Boolean rule. 
This canonical property is useful because it preserves the original CA rule on binary states and gives an explicit polynomial update function. 
Several studies have exploited this tractability, for example in analyses of convergence and aperiodicity for fuzzy rule 90 \cite{Flocchini2000}, fuzzy rule 110 dynamics \cite{Mingarelli2003}, classification of fuzzy ECAs \cite{Mingarelli2010}, rule 184 fuzzy traffic-flow dynamics \cite{Higashi2021}, and explicit asymptotic solutions for fuzzy rule 38 \cite{Yamamoto2023Asymptotic}.

At the same time, this multi-affine structure can be restrictive. Since FDNF interpolates the Boolean vertex values smoothly inside the unit cube, it can attenuate the high-contrast structures responsible for characteristic CA patterns. 
For representative rules such as those considered below, the ordinary FDNF dynamics may therefore become much smoother than the corresponding Boolean ECA dynamics, sometimes approaching homogeneous, monotone, checkerboard-like, or otherwise regular asymptotic profiles \cite{Flocchini2000,Higashi2021,Yamamoto2023Asymptotic}. 
This creates a gap between two desirable requirements: a fuzzification should preserve the original Boolean rule on binary states, but it should also allow sufficiently flexible continuous-state dynamics to retain or generate nontrivial patterns.

A related but different direction is provided by application-oriented CA and FCA models. In image analysis and classification, CA- and fuzzy-model-based methods have been used for denoising, edge detection, and feature extraction \cite{Nayak2014,Sadjadi2022,Tangsakul2023}. 
Other examples include urban-growth modeling with transition rules optimized by genetic fuzzy systems \cite{Foroutan2022} and fuzzy cellular-automaton traffic-flow models with bottlenecks \cite{Nishida2022}. 
These studies demonstrate the modeling flexibility of fuzzy or continuous-state cellular automata. However, when the fuzzy transition rule is designed empirically or optimized for a specific task, it may no longer be systematically related to a prescribed Boolean CA rule. 
Consequently, one may lose structural advantages of FDNF, such as its explicit rule-preserving construction and its compatibility with analytical methods.

The aim of this paper is to develop a systematic framework that lies between these two directions. 
Starting from the FDNF polynomial \(f_k\) of a Boolean ECA rule, we compose it with endpoint-preserving transformations:
\[
\widetilde f_k^{g,u,v,w}(x,y,z) = g\left(f_k(u(x),v(y),w(z))\right),
\]
where \(g,u,v,w:[0,1]\to[0,1]\) satisfy
\[
g(0)=u(0)=v(0)=w(0)=0, \qquad g(1)=u(1)=v(1)=w(1)=1.
\]
This construction preserves the original Boolean rule on \(\{0,1\}^3\), and the ordinary FDNF is recovered when all four transformations are the identity. 
By choosing different classes of endpoint-preserving transformations, including continuous, Lipschitz, piecewise-linear, discontinuous, non-monotone, or expanding maps, one obtains a broad family of rule-preserving fuzzy extensions of the same Boolean rule.

This framework has two purposes. 
The first is constructive: it provides a rule-preserving way to generate many fuzzy cellular automata from a single Boolean CA without abandoning the FDNF structure. 
The second is dynamical:
it allows one to tune and analyze how the fuzzification affects pattern formation and temporal evolution. For example, a threshold-like fuzzification function can promote ECA-like binary patterns from non-binary initial data, while a one-parameter family connecting this threshold-like case to ordinary FDNF reveals how patterns are gradually smoothed out. Discontinuous or non-monotone fuzzification functions can produce pattern regimes that are not accessible through ordinary FDNF.

The generalized FDNF framework also connects fuzzy cellular automata with continuous-state nonlinear dynamical systems. Even in the minimal three-cell case, a Boolean ECA has only finitely many states, so every orbit is eventually periodic. 
In contrast, the corresponding generalized fuzzy ECA defines a map on the unit cube \([0,1]^3\). With suitable expanding fuzzification functions, such maps can exhibit locally stable periodic cycles, expanding invariant line-set dynamics, and finely mixed finite-resolution class structures. 
Thus, the proposed construction is not only a generalization of FDNF, but also a simple mechanism for generating nonlinear dynamics from Boolean cellular-automaton rules.

The main contributions of this paper are as follows. First, we formulate FDNF as the canonical multi-affine extension of an ECA rule and introduce its generalization through endpoint-preserving fuzzification functions. 
Second, we show that the proposed construction contains ordinary FDNF as a special case and provides a systematic class of fuzzy extensions preserving the original Boolean rule at binary states. 
Third, we demonstrate, through representative rules, how the choice of fuzzification function affects pattern formation, including the transition from ECA-like patterns to FDNF-like smoothing and the creation of gap-induced patterns. 
Fourth, we introduce finite-resolution quantitative indicators, including contrast, fuzziness, and a participation-type support exponent, to summarize amplitude-support changes in these pattern regimes. 
Finally, we analyze minimal three-cell generalized fuzzy ECAs and
show that expanding endpoint-preserving transformations can generate locally stable periodic cycles, expanding invariant line-set dynamics, and finely mixed finite-resolution slice structures.

The rest of the paper is organized as follows. Section~2 reviews ECA rules and FDNF fuzzification, emphasizing the interpretation of FDNF as a multi-affine extension. 
Section~3 introduces endpoint-preserving fuzzification functions and defines the generalized FDNF construction.
Section~4 studies pattern formation under several classes of fuzzification functions, including parameter families connecting ECA-like and FDNF-like behavior. 
Section~5 introduces an effective support exponent and applies it to parameter-dependent pattern changes generated by the deformation family \(q_a\) and the discontinuous gap family \(r_a\). Section~6 investigates three-cell generalized fuzzy ECAs, with particular attention to locally stable periodic cycles, expanding invariant line-set dynamics, and finite-resolution class structures.
Section~7 concludes the paper and discusses further directions. Additional numerical protocols, resolution checks, and an illustrative convex-mixing extension beyond the single-rule GFDNF setting are provided in the appendices.

%%%%%%%%%%%%%%%%%%%%%%%%%%%%%%%%%%%%%%%%%%%%%%%%%%%%%%%%%%%%%%
%   section 2 
%%%%%%%%%%%%%%%%%%%%%%%%%%%%%%
\section{FDNF as the canonical multi-affine fuzzification}
\label{sec:fdnf}

We first recall elementary cellular automata and their fuzzification by fuzzy disjunctive normal form (FDNF). 
Let \(s_j^t\in\{0,1\}\) denote the state of the \(j\)-th cell at time \(t\). 
An elementary cellular automaton (ECA) is a one-dimensional Boolean cellular automaton whose local rule depends on the left, center, and right neighboring states. 
For rule \(k\), the time evolution is given by
\begin{equation}
s_j^{t+1}=F_k(s_{j-1}^t,s_j^t,s_{j+1}^t),
\label{eq:eca_update}
\end{equation}
where
\[
F_k:\{0,1\}^3\to\{0,1\}.
\]
The rule \(F_k\) is determined by the eight Boolean values
\[
a_i\in\{0,1\},\qquad i=0,\ldots,7,
\]
ordered as
\[
\begin{array}{c|cccccccc}
x\,y\,z &000&001&010&011&100&101&110&111\\
\hline
F_k(x,y,z)&a_0&a_1&a_2&a_3&a_4&a_5&a_6&a_7 .
\end{array}
\]
The rule number is defined by
\begin{equation}
k=\sum_{i=0}^{7} a_i2^i .
\label{eq:rule_number}
\end{equation}
For example, rule 184 is given by
\[
(a_0,a_1,a_2,a_3,a_4,a_5,a_6,a_7)
=
(0,0,0,1,1,1,0,1),
\]
and is known as a basic ECA model of traffic flow\cite{Nagel1992,Biham1992}.

We now define the FDNF fuzzification of \(F_k\). Put
\[
\beta_0(r)=1-r,\qquad \beta_1(r)=r
\]
for \(r\in[0,1]\). For \((x_1,x_2,x_3)=(x,y,z)\), the FDNF polynomial associated
with \(F_k\) is
\begin{equation}
f_k(x,y,z)
=
\sum_{\varepsilon_1,\varepsilon_2,\varepsilon_3\in\{0,1\}}
F_k(\varepsilon_1,\varepsilon_2,\varepsilon_3)
\prod_{m=1}^{3}\beta_{\varepsilon_m}(x_m).
\label{eq:fdnf_basis}
\end{equation}
Equivalently, in terms of the coefficients \(a_i\), this is written as
\begin{equation}
\begin{split}
f_k(x,y,z)
=&a_0(1-x)(1-y)(1-z)
+a_1(1-x)(1-y)z\\
&+a_2(1-x)y(1-z)
+a_3(1-x)yz 
+a_4x(1-y)(1-z)\\
&+a_5x(1-y)z
+a_6xy(1-z)
+a_7xyz .
\label{eq:fdnf}
\end{split}
\end{equation}
This is the standard FDNF rule obtained by replacing the Boolean minterms in the
disjunctive normal form by their fuzzy counterparts.

The following proposition gives the mathematical characterization of FDNF that will be used throughout this paper.

\begin{proposition}%[FDNF as the unique multi-affine extension]
\label{prop:fdnf_unique}
$\quad$\\
\indent
Let \(F_k:\{0,1\}^3\to\{0,1\}\) be an ECA local rule. Then the FDNF polynomial \(f_k\) defined by \eqref{eq:fdnf_basis} is the unique function \(h:[0,1]^3\to\mathbb{R}\) satisfying the following two properties:
\begin{enumerate}
\item \(h\) is affine in each variable separately;
\item \(h(\varepsilon_1,\varepsilon_2,\varepsilon_3)
=
F_k(\varepsilon_1,\varepsilon_2,\varepsilon_3)\) \ for all \ 
\((\varepsilon_1,\varepsilon_2,\varepsilon_3)\in\{0,1\}^3\).
\end{enumerate}
Moreover, \(
0\le f_k(x,y,z)\le 1\)
for all  \( (x,y,z)\in[0,1]^3.
\)
\end{proposition}

\begin{proof}
The definition of $f_k$ clearly gives Properties 1 and 2. 
Since $f_k$ is multi-affine, its maximum and minimum over the compact cube \([0,1]^3\) are attained at vertices. 
Since all vertex values are either 0 or 1, it follows that \(0\le f_k \le 1\). Conversely, any multi-affine function on \([0,1]^3\) is represented by its tensor-product multilinear interpolation formula.
\[
h(x,y,z)
=
\sum_{\varepsilon_1,\varepsilon_2,\varepsilon_3\in\{0,1\}}
h(\varepsilon_1,\varepsilon_2,\varepsilon_3)
\prod_{m=1}^{3}\beta_{\varepsilon_m}(x_m).
\]
If \(h\) agrees with \(F_k\) on the eight vertices of the unit cube, this expression coincides with \eqref{eq:fdnf_basis}. Therefore \(h=f_k\), proving uniqueness.
\end{proof}

Proposition \ref{prop:fdnf_unique} shows that FDNF is the canonical interpolation of the Boolean rule among all multi-affine extensions. 
At the same time, the proposition also indicates a limitation:
FDNF is unique only within the multi-affine class. 
There are many other endpoint-preserving extensions of the same Boolean rule if one allows nonlinear, piecewise-linear, discontinuous, or non-monotone functions. This observation is the starting point of the generalized construction introduced in Section~\ref{sec:gfdnf}.

The fuzzy elementary cellular automaton (FECA) obtained by FDNF is defined with the FDNF polynomial \eqref{eq:fdnf_basis} as a dynamical system by allowing the cell states to take values in \([0,1]\). Namely, for \(x_j^t\in[0,1]\), we set
\begin{equation}
x_j^{t+1}=f_k(x_{j-1}^t,x_j^t,x_{j+1}^t).
\label{eq:feca_update}
\end{equation}
Since \(f_k\) agrees with \(F_k\) on \(\{0,1\}^3\), the FECA exactly reproduces the original ECA whenever the initial configuration is binary.

\begin{figure}[htbp]
  % 左側の図
  \begin{minipage}[b]{0.48\textwidth}
    \centering
    \includegraphics[width=\linewidth]{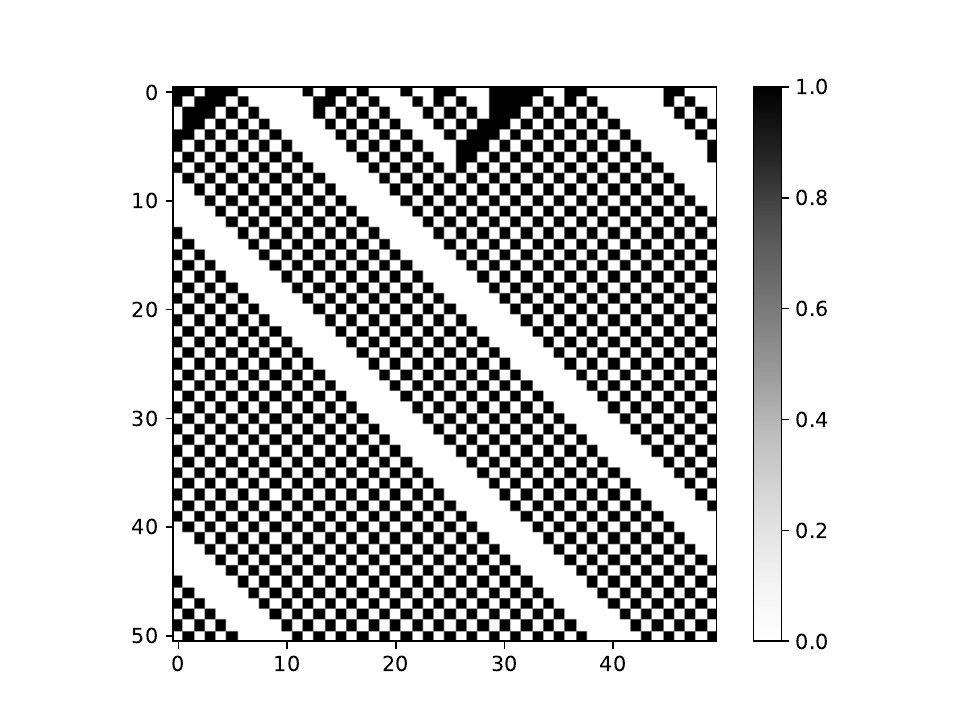}
\caption{A time evolution pattern of the Boolean rule 184 ECA.}
\label{fig:rule184_eca}
  \end{minipage}
  \hfill % 図と図の間のスペースを自動で空ける
  % 右側の図
  \begin{minipage}[b]{0.48\textwidth}
    \centering
    \includegraphics[width=\linewidth]{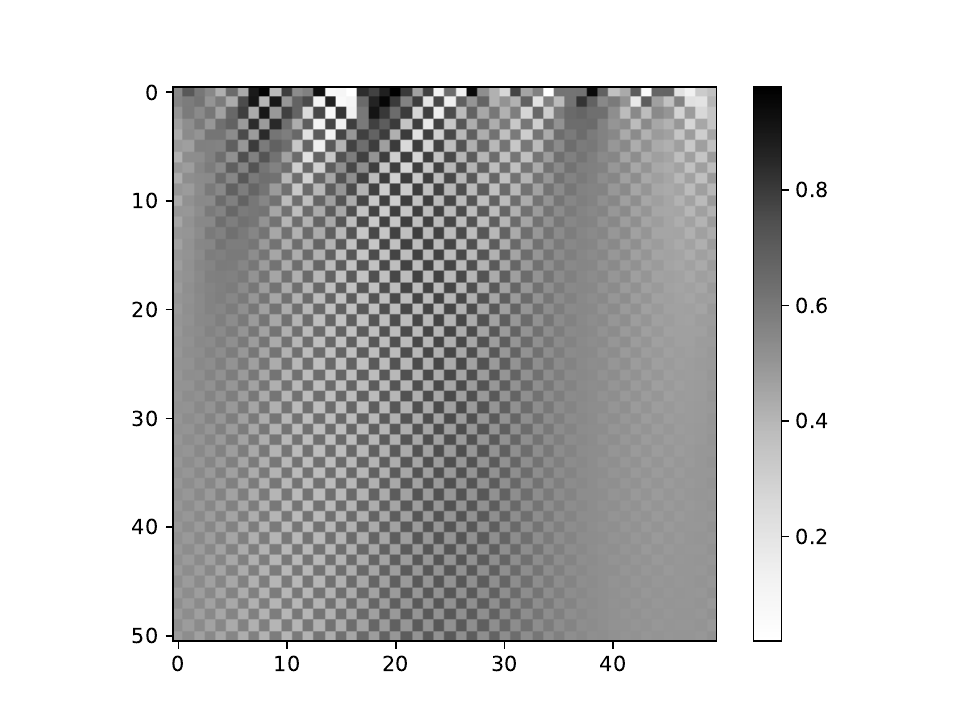}
\caption{A time evolution pattern of the rule 184 FECA obtained by FDNF.
%The system size is 50, the initial values are chosen randomly from \([0,1]\), and a periodic boundary condition is imposed.
}
\label{fig:rule184_fdnf}
  \end{minipage}
\end{figure}

We next illustrate how the multi-affine nature of FDNF affects pattern formation.
For rule 184, Eq.~\eqref{eq:fdnf} gives
\begin{align}
f_{184}(x,y,z)
&=(1-x)yz+x(1-y)(1-z)+x(1-y)z+xyz \notag\\
&=x-xy+yz .
\label{eq:f184}
\end{align}
Figure~\ref{fig:rule184_eca} shows a typical time evolution of the Boolean rule 184 ECA, while Figure~\ref{fig:rule184_fdnf} shows the corresponding FDNF-based FECA from random initial values in \([0,1]\). In contrast to the high-contrast binary pattern of the ECA, the FDNF dynamics produces a much smoother pattern. For rule 184, this smoothing is consistent with the known asymptotic behavior of the FDNF-based fuzzy rule: depending on the parity of the system size, the solution approaches either a monotone pattern or a checkerboard-type pattern.

Similar smoothing is observed for other representative rules. For example,
\[
f_{90}(x,y,z)=x+z-2xz,
\label{eq:f90}
\]
and
\[
f_{30}(x,y,z)
=
2xyz-2xy-2xz+x-yz+y+z.
\label{eq:f30}
\]

%%%%%%

\begin{figure}[htb]
\centering
\includegraphics[width=0.80\textwidth]{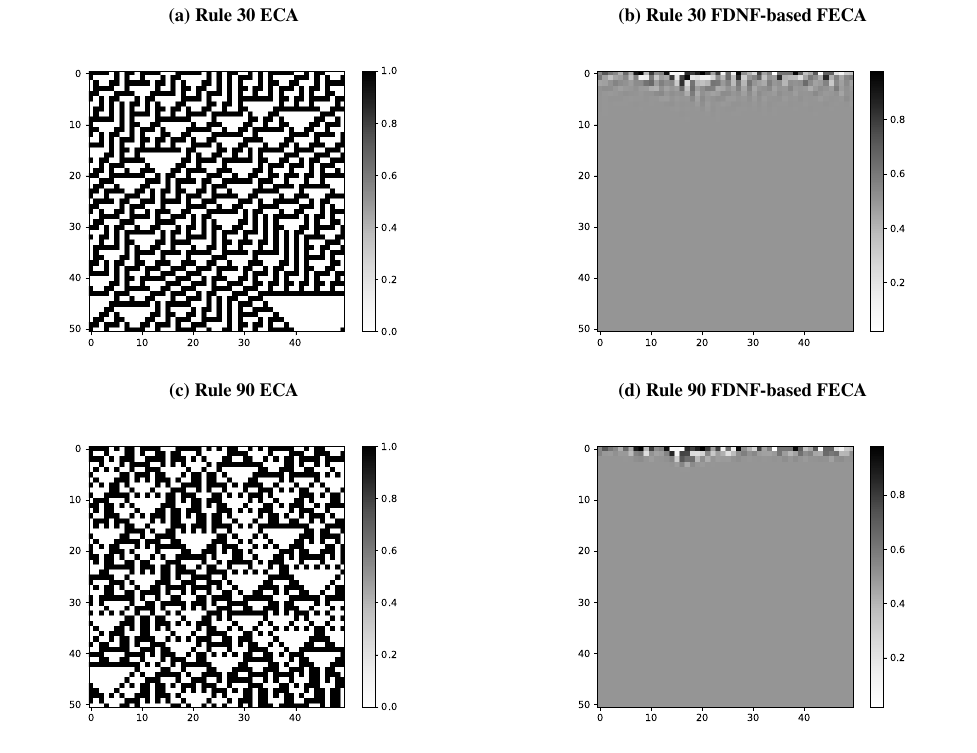}
\caption{Comparison between Boolean ECA patterns and FDNF-based FECA patterns.
Upper panels: rule 30. Lower panels: rule 90.}
\label{fig:rule30_rule90_fdnf}
\end{figure}

Figure~\ref{fig:rule30_rule90_fdnf} compares the Boolean ECA patterns and the corresponding FDNF-based FECA patterns for rules 30 and 90. The original ECAs generate complex binary spatiotemporal structures, whereas the FDNF versions rapidly lose the high-contrast cellular-automaton patterns and tend toward nearly homogeneous or regular profiles.

These examples do not imply that all FDNF-based fuzzy cellular automata are dynamically trivial. Rather, they show that the canonical multi-affine extension can suppress characteristic CA patterns for important representative rules. Thus, FDNF has both a strength and a limitation: it is mathematically canonical and preserves the Boolean rule at the vertices, but its multi-affine structure restricts the range of continuous-state dynamics. The next section generalizes this construction while retaining the endpoint-preserving property.

\FloatBarrier

%%%%%%%%%%%%%%%%%%%%%%%%%%%%%%%%%%%%%%%%%%%%%%%%%%%%%%%%%
%
%     section 3  
%%%%%%%%%%%%%%%%%%%%%%%%%%%%%%%%%%%%%%%%%%%%%%%%%%%%%%
\section{Generalized FDNF by endpoint-preserving fuzzification functions}
\label{sec:gfdnf}

% If the package placeins is loaded, \FloatBarrier prevents the table and
% figure in this section from floating into the next section. If placeins is
% not loaded, the following fallback keeps the source compilable.
\providecommand{\FloatBarrier}{}

In the previous section, the FDNF polynomial \(f_k\) was characterized as the canonical multi-affine extension of the Boolean ECA rule \(F_k\). 
In this section, we introduce a systematic generalization of FDNF that keeps the endpoint-preserving property but allows a much wider class of fuzzy dynamics.

We first define the broad class of endpoint-preserving fuzzy extensions of an ECA rule. Let
\[
F_k:\{0,1\}^3\to\{0,1\}
\]
be the local rule of the rule \(k\) ECA.

\begin{definition}[Endpoint-preserving fuzzy extension]
\label{def:endpoint_extension}
An endpoint-preserving fuzzy extension of the ECA rule \(F_k\) is a map
\(
\Phi_k:[0,1]^3\to[0,1]
\)
satisfying
\(
\Phi_k(\varepsilon_1,\varepsilon_2,\varepsilon_3)
=
F_k(\varepsilon_1,\varepsilon_2,\varepsilon_3)
\)
for all
\(
(\varepsilon_1,\varepsilon_2,\varepsilon_3)\in\{0,1\}^3.
\)

The corresponding one-dimensional, radius-one fuzzy cellular automaton is defined by
\begin{equation}
 x_j^{t+1}
 =
 \Phi_k(x_{j-1}^t,x_j^t,x_{j+1}^t),
 \qquad x_j^t\in[0,1].
\label{eq:endpoint_fca_update}
\end{equation}
We call such a system a generalized fuzzy elementary cellular automaton, or GFECA, associated with rule \(k\).
\end{definition}

Equivalently, the set of all endpoint-preserving fuzzy extensions of rule \(k\) is
\begin{equation}
\mathcal E_k
=
\left\{
\Phi:[0,1]^3\to[0,1]
\mid
\Phi|_{\{0,1\}^3}=F_k \right\}.
\label{eq:Ek}
\end{equation}
The ordinary FDNF polynomial \(f_k\) belongs to \(\mathcal E_k\). However, \(\mathcal E_k\) is much larger than the multi-affine class. The purpose of this paper is not to study all elements of \(\mathcal E_k\), but to introduce a structured and tractable subclass obtained by composing the FDNF polynomial with endpoint-preserving one-dimensional functions.

\begin{definition}[Endpoint-preserving fuzzification functions]
\label{def:omega}
Let
\begin{equation}
\Omega =
\left\{ q:[0,1]\to[0,1] \mid q(0)=0,\ q(1)=1 \right\}.
\label{eq:omega}
\end{equation}
An element \(q\in\Omega\) is called an endpoint-preserving fuzzification function.
\end{definition}

We use the term fuzzification function in a broad mathematical sense. In the present paper it means an endpoint-preserving state transformation on \([0,1]\), not necessarily a monotone membership transformation. 
Thus, discontinuous, non-monotone, and expanding maps may be included when they are useful for constructing continuous-state dynamics from a Boolean rule.

The endpoint conditions in \eqref{eq:omega} are the essential requirement. The function \(q\) need not be linear, monotone, continuous, or differentiable. 
This flexibility allows us to construct fuzzy rules that are still consistent with the original Boolean rule at binary states, but whose behavior inside the unit interval can be chosen according to the desired dynamics.

\begin{definition}[Generalized FDNF rule]
\label{def:gfdnf}
Let \(g,u,v,w\in\Omega\). The generalized FDNF rule associated with the ECA rule \(F_k\) is defined by
\begin{equation}
\widetilde f_k^{g,u,v,w}(x,y,z)
=
g\left(f_k(u(x),v(y),w(z))\right), \qquad (x,y,z)\in[0,1]^3.
\label{eq:gfdnf}
\end{equation}
When no confusion is possible, we simply write \(\widetilde f_k\) instead of \(\widetilde f_k^{g,u,v,w}\). The corresponding GFECA is given by
\begin{equation}
x_j^{t+1}
=
\widetilde f_k^{g,u,v,w}
(x_{j-1}^t,x_j^t,x_{j+1}^t).
\label{eq:gfdnf_update}
\end{equation}
We use the abbreviation GFDNF for rules of the form \eqref{eq:gfdnf}, and GFECA for the corresponding cellular automata defined by \eqref{eq:gfdnf_update}.
\end{definition}

The following proposition is easily proved from the definition of the fuzification functions, but it is the central rule-preservation property of the proposed construction.

\begin{proposition}%[Endpoint preservation of generalized FDNF]
\label{prop:gfdnf_endpoint}
\hfill\break
For any \(g,u,v,w\in\Omega\), the generalized FDNF rule \(\widetilde f_k^{g,u,v,w}\) belongs to \(\mathcal E_k\). 
In particular, \(\widetilde f_k^{g,u,v,w}\) defines a GFECA associated with rule \(k\).
\end{proposition}

An immediate consequence is that binary configurations are invariant under the generalized FDNF dynamics.

\begin{corollary}%[Recovery of the original ECA on binary configurations]
\label{cor:binary_recovery}
Suppose that the initial configuration satisfies
\(
x_j^0\in\{0,1\}
\)
for all \(j\). Then the GFECA defined by \eqref{eq:gfdnf_update} evolves exactly as the rule \(k\) ECA. In particular,
\(
x_j^t\in\{0,1\}
\)
for all \(j\) and all \(t\ge 0\), and
\(
x_j^{t+1}
=
F_k(x_{j-1}^t,x_j^t,x_{j+1}^t).
\)
\end{corollary}

\begin{proof}
If \(x_{j-1}^t,x_j^t,x_{j+1}^t\in\{0,1\}\), then Proposition \ref{prop:gfdnf_endpoint} gives
\[
\widetilde f_k^{g,u,v,w}(x_{j-1}^t,x_j^t,x_{j+1}^t)
=
F_k(x_{j-1}^t,x_j^t,x_{j+1}^t) \in\{0,1\}.
\]
The assertion follows by induction on \(t\).
\end{proof}

The class of generalized FDNF rules can be written as
\[
\mathcal G_k(\Omega)
=
\left\{
g\circ f_k\circ(u,v,w)
\mid
g,u,v,w\in\Omega
\right\},
\]
where
\(
(u,v,w)(x,y,z)=(u(x),v(y),w(z)).
\)
Thus
\(
\ \mathcal G_k(\Omega)\subset \mathcal E_k.
\)

This inclusion can be strict. For example, for rule \(0\) we have \(f_0\equiv 0\), and hence every rule in \(\mathcal G_0(\Omega)\) is identically zero because \(g(0)=0\). 
On the other hand, the map
\[
\Phi(x,y,z)=x(1-x)y(1-y)z(1-z)
\]
belongs to \(\mathcal E_0\), since it vanishes on all Boolean vertices, but it is not identically zero. Thus \(\mathcal G_0(\Omega)\subsetneq\mathcal E_0\).
For a general rule \(k\), the point of the construction is not to characterize all of \(\mathcal E_k\), but to provide a systematic FDNF-based subclass rather than an arbitrary collection of fuzzy transition rules.

For later use, we also introduce several subclasses of \(\Omega\). Let
\[
\Omega_C = \Omega\cap C([0,1])
\]
be the class of continuous endpoint-preserving fuzzification functions. For \(L>0\), let
\[
\Omega_{\mathrm{Lip}}(L)
=
\left\{
q\in\Omega
\mid
|q(x)-q(y)|\le L|x-y|
\quad\text{for all }x,y\in[0,1]
\right\}.
\]
We also use the class \(\Omega_{\mathrm{PL}}\) of functions \(q\in\Omega\) that are linear on finitely many subintervals of \([0,1]\), allowing jump discontinuities at the break points. More precisely, such a function is affine on each open interval of a finite partition of \([0,1]\), and the one-sided values at the break points are allowed to differ. Non-monotone and expanding maps are also allowed when they belong to \(\Omega\).

The regularity of the generalized FDNF rule is inherited from the fuzzification functions.

\begin{proposition}%[Regularity inheritance]
\label{prop:regularity}
Let \(g,u,v,w\in\Omega\). Then the following statements hold.
\begin{enumerate}
\item If \(g,u,v,w\in\Omega_C\), then \(\widetilde f_k^{g,u,v,w}\) is continuous.
\item If \(g,u,v,w\) are Lipschitz continuous, then \(\widetilde f_k^{g,u,v,w}\) is Lipschitz continuous.
\item If \(g,u,v,w\) are piecewise linear in the above sense, with finitely many pieces, then \(\widetilde f_k^{g,u,v,w}\) is piecewise polynomial.
\end{enumerate}
\end{proposition}

\begin{proof}
The first assertion follows from the continuity of compositions. For the second, let \(L_g,L_u,L_v,L_w\) be Lipschitz constants of \(g,u,v,w\), respectively, and define
\[
M_x(k)=\sup_{[0,1]^3}|\partial_x f_k|,
\quad
M_y(k)=\sup_{[0,1]^3}|\partial_y f_k|,
\quad
M_z(k)=\sup_{[0,1]^3}|\partial_z f_k|.
\]
Since \(f_k\) is a polynomial, these quantities are finite. By the mean value theorem,
\[
\operatorname{Lip}_\infty
\left( \widetilde f_k^{g,u,v,w} \right)
\le
L_g\left( M_x(k)L_u+M_y(k)L_v+M_z(k)L_w \right),
\]
where \(\operatorname{Lip}_\infty\) denotes the Lipschitz constant with respect to the maximum norm on \([0,1]^3\).

For the third assertion, the break points of \(u,v,w\) divide the cube \([0,1]^3\) into finitely many rectangular cells. On each such cell, \(f_k(u(x),v(y),w(z))\) is a polynomial. The break points of \(g\) further divide each cell by the inverse images of finitely many intervals under this polynomial. 
Hence the composition is polynomial on each element of a finite
semialgebraic partition. 
This is the sense in which \(\widetilde f_k^{g,u,v,w}\) is piecewise polynomial.
\end{proof}

We now summarize several important special cases of the generalized FDNF construction in Table~\ref{tab:fuzzification_classes}. 
These cases explain how the choice of fuzzification functions changes the resulting fuzzy dynamics.

\begin{table}[!htbp]
\centering
\caption{Typical choices of fuzzification functions in the generalized FDNF framework.}
\label{tab:fuzzification_classes}
\small
\setlength{\tabcolsep}{5pt}
\renewcommand{\arraystretch}{1.18}
\begin{tabularx}{\textwidth}{@{}p{0.34\textwidth}Y@{}}
\toprule
Choice of fuzzification functions & Interpretation and dynamical role \\
\midrule
\(g=u=v=w=\mathrm{id}\)
& Ordinary FDNF is recovered. This is the canonical multi-affine fuzzification of the Boolean ECA rule. \\

Threshold-like \(u,v,w\)
& The arguments of \(f_k\) are pushed toward binary values, leading to ECA-like pattern recovery from fuzzy initial data. \\

Piecewise-linear deformation
& The fuzzification can interpolate between threshold-like ECA behavior and ordinary FDNF, allowing controlled smoothing of patterns. \\

Discontinuous map with a gap
& A jump discontinuity can separate nearby fuzzy states into different branches and generate pattern regimes not obtained by ordinary FDNF. \\

Non-monotone or expanding map
& Expanding or folding behavior can produce low-dimensional nonlinear dynamics, including inherited expanding dynamics, locally stable cycles, and finite-resolution class structures. \\
\bottomrule
\end{tabularx}
\end{table}

\begin{figure}[!htbp]
\centering
\includegraphics[width=0.75\textwidth]{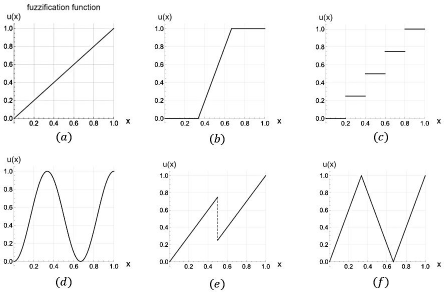}
\caption{Examples of endpoint-preserving fuzzification functions.
(a) Identity map, which recovers ordinary FDNF;
(b) continuous threshold-like map;
(c) step-like discontinuous map;
(d) non-monotone map;
(e) discontinuous gap map;
(f) non-monotone expanding map.
These examples illustrate how endpoint-preserving transformations can interpolate between ECA-like and FDNF-like behavior or generate new continuous-state dynamics.}
\label{fig:fuzzification_functions}
\end{figure}

Figure~\ref{fig:fuzzification_functions} shows representative elements of \(\Omega\). The identity map recovers ordinary FDNF, threshold-like maps promote ECA-like binary behavior, discontinuous maps can separate nearby states into distinct branches, and non-monotone expanding maps provide a source of
nonlinear temporal dynamics. 
The following sections study these cases through concrete ECA rules.

\FloatBarrier

%%%%%%%%%%%%%%%%%%%%%%%%%%%%%%%%%%%%%%%%%%%%%%%%%%%%%%%%%%%%%%%%%
% section 4 
%%%%%%%%%%%%%%%%%%%%%%%%%%%%%%%%%%%%%%%%%%%%%%%%%%%%%%%%%%%%%%%%%
\section{Pattern preservation and loss under generalized FDNF}
\label{sec:pattern_preservation}

The generalized FDNF construction introduced in Section~\ref{sec:gfdnf} contains ordinary FDNF as a special case, but it also allows us to move away from the multi-affine FDNF dynamics by changing the endpoint-preserving fuzzification functions. 
The purpose of this section is to illustrate, through representative ECA rules, how different endpoint-preserving transformations can promote ECA-like pattern recovery, interpolate toward FDNF-like smoothing, or induce gap-generated pattern regimes.

Throughout this section, we consider a one-dimensional lattice of size \(N\) with periodic boundary conditions. Unless otherwise stated, we take
\[
g(x)=x,\qquad u(x)=v(x)=w(x)=q(x),
\]
so that the update rule is
\[
x^{t+1}_j
=
f_k\left(q(x^t_{j-1}),q(x^t_j),q(x^t_{j+1})\right),
\qquad x^t_j\in[0,1].
\]
Here \(f_k\) is the FDNF polynomial of the ECA rule \(F_k\). 
Since \(q\in\Omega\), the resulting rule still agrees with the original ECA rule on binary configurations.

\subsection{Threshold-like fuzzification and recovery of ECA-like patterns}

We first consider a continuous threshold-like fuzzification function:
\begin{equation}
q_0(x)=
\begin{cases}
0, & 0\le x\le \dfrac13,\\[1mm]
3x-1, & \dfrac13 < x\le \dfrac23,\\[1mm]
1, & \dfrac23 < x\le 1.
\end{cases}
\label{eq:q0}
\end{equation}
This is the function shown in Fig.~\ref{fig:fuzzification_functions}(b).
The plateaus near 0 and 1 push fuzzy states toward binary values, while the middle linear branch keeps the map continuous.

\begin{figure}[htb]
\centering
\includegraphics[width=0.75\textwidth]{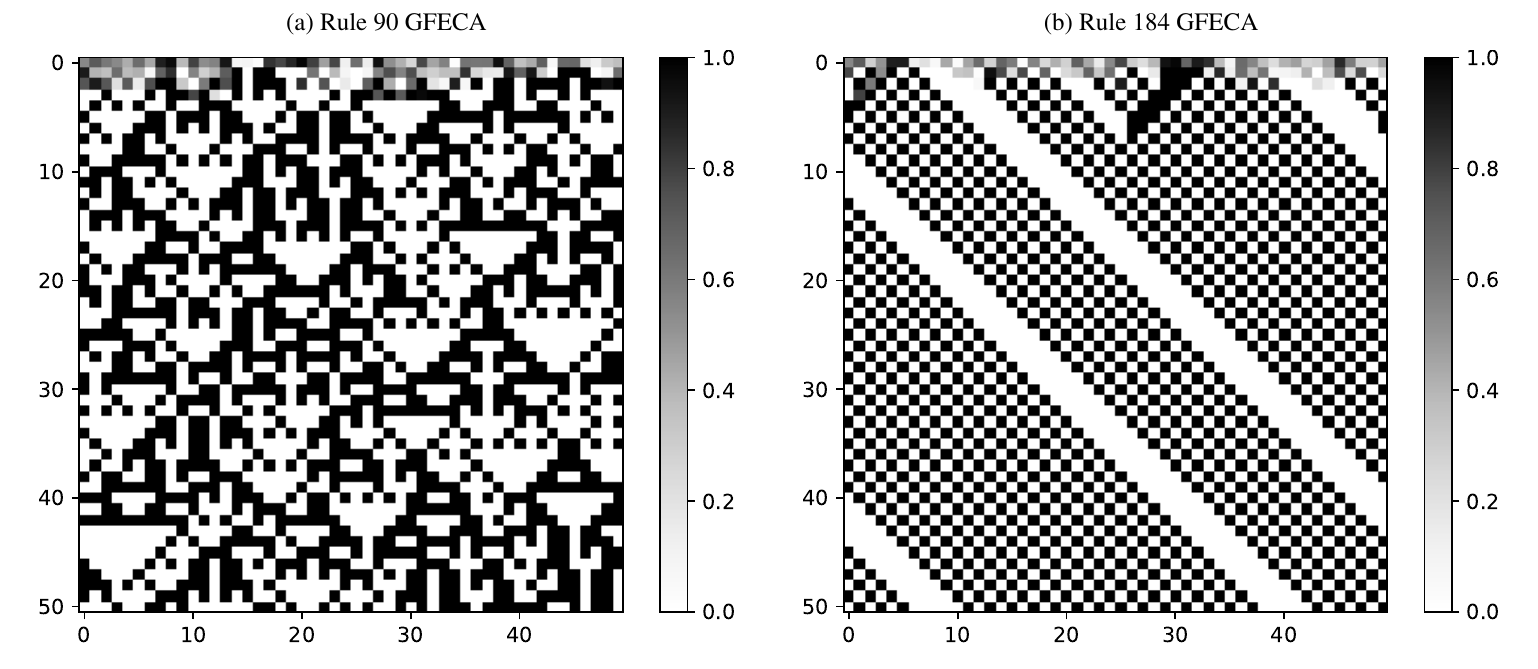}
\caption{Time-evolution patterns generated by the threshold-like fuzzification function \(q_0\) in Eq.~\eqref{eq:q0}. Left: rule 90 GFECA. Right: rule 184 GFECA. 
Although the initial values are chosen randomly from \([0,1]\), the patterns become close to the corresponding binary ECA patterns.}
\label{fig:threshold_patterns}
\end{figure}

Using \(q_0\) in the above update rule, we obtain GFECAs that often recover high-contrast patterns close to the corresponding binary ECA patterns from non-binary initial data. 
Figure~\ref{fig:threshold_patterns} shows typical time-evolution patterns for rules 90 and 184. In both cases, the initial values are chosen randomly from \([0,1]\), but the resulting patterns become close to the corresponding binary ECA patterns.

The exact recovery mechanism is clearest in the limiting case of a step-type fuzzification function. For \(0<\theta<1\), define
\[
Q_\theta(x)=
\begin{cases}
0, & 0\le x\le \theta,\\
1, & \theta < x\le 1.
\end{cases}
\]
Consider the GFECA with \(g=\mathrm{id}\) and \(u=v=w=Q_\theta\). 
For any \(x,y,z\in[0,1]\), the values \(Q_\theta(x),Q_\theta(y),Q_\theta(z)\) are binary. Since the FDNF polynomial \(f_k\) agrees with the Boolean rule \(F_k\) on the Boolean vertices, we have
\[
f_k\bigl(Q_\theta(x),Q_\theta(y),Q_\theta(z)\bigr)
=
F_k\bigl(Q_\theta(x),Q_\theta(y),Q_\theta(z)\bigr).
\]
Let the thresholded initial configuration be
\[
s_j^0=Q_\theta(x_j^0),
\]
and let \(s_j^t\) evolve according to the Boolean ECA rule
\[
s_j^{t+1}=F_k(s_{j-1}^t,s_j^t,s_{j+1}^t).
\]
Then the step-fuzzified GFECA satisfies
\[
x_j^t=s_j^t \qquad (t\ge 1).
\]
Thus the step fuzzification exactly reduces the continuous-state dynamics to the Boolean ECA dynamics generated from the thresholded initial configuration, after the first update. 
The continuous threshold-like map \(q_0\) in Eq.~\eqref{eq:q0} is a regularized version of this exact-recovery limit: it preserves the endpoint rule consistency of FDNF while promoting, rather than guaranteeing, ECA-like binary pattern formation from non-binary initial data.
%%%%%%%%%%%%%%%%%%%%%
%%%%%%%%%%%%%%%%%%%%%
\subsection{A deformation from ECA-like behavior to ordinary FDNF}
\label{subsec:qa_deformation}

We next introduce a one-parameter family of fuzzification functions that connects the threshold-like map \(q_0\) to the identity map. 
For \(0\le a\le 1\), define
\begin{equation}
q_a(x)=
\begin{cases}
a x, & 0\le x\le \dfrac{1}{3-a},\\[2mm]
3x-1, & \dfrac{1}{3-a}<x\le \dfrac{2-a}{3-a},\\[2mm]
a x+1-a, & \dfrac{2-a}{3-a}<x\le 1.
\end{cases}
\label{eq:qa}
\end{equation}
Then \(q_a\in\Omega\) for all \(0\le a\le 1\), and
\(a=0\) corresponds to a threshold-like ECA-recovery endpoint, whereas \(a=1\) recovers ordinary FDNF. The parameter \(a\) is also the one-sided derivative of \(q_a\) at the endpoints:
\(
q_a'(0)=q_a'(1)=a.
\)
Therefore, increasing \(a\) weakens the contraction toward binary states and moves the dynamics toward the FDNF endpoint.

\begin{figure}[htb]
\centering
\includegraphics[width=0.5\textwidth]{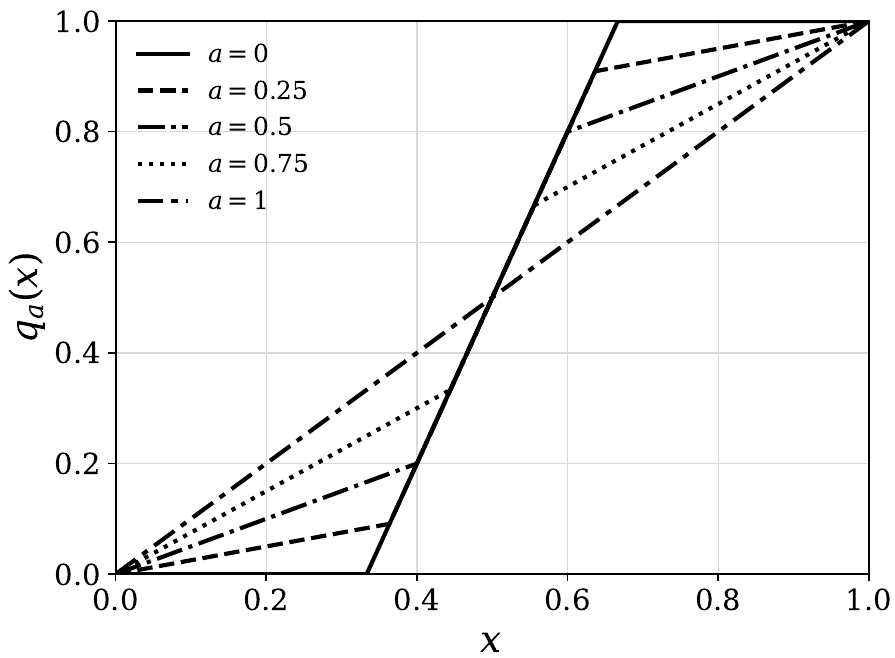}
\caption{The one-parameter family of fuzzification functions \(q_a\).
The case \(a=0\) gives the threshold-like map \(q_0\), while \(a=1\)
corresponds to the identity map and hence recovers ordinary FDNF.}
\label{fig:qa_family}
\end{figure}

\begin{figure}[hbt]
\centering
\includegraphics[width=1.0\textwidth]{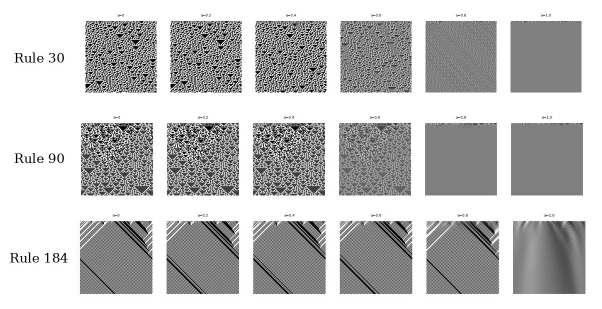}
\caption{
Time-evolution patterns of the GFECAs for rules 30, 90, and 184 using the one-parameter fuzzification function \(q_a\). The system size is \(N=100\), periodic boundary conditions are imposed, and the initial values are chosen randomly from \([0,1]\). The columns correspond to \(a=0,0.2,0.4,0.6,0.8,1.0\). For each rule, the same random initial condition was used for all values of \(a\).
}
\label{fig:qa_rule30_rule90_rule184}
\end{figure}

Figure~\ref{fig:qa_rule30_rule90_rule184} shows time-evolution patterns of the GFECAs for rules 30, 90, and 184, respectively, using \(q_a\). The simulations are performed with periodic boundary conditions from random initial values in \([0,1]\). For each rule, the same initial condition is used for all displayed values of \(a\), so that the change across columns reflects the effect of the fuzzification parameter. The values of the parameter are
\[
a=0,\ 0.2,\ 0.4,\ 0.6,\ 0.8,\ 1.
\]
For \(a=0\), the patterns remain close to binary ECA-like patterns. As \(a\) increases toward 1, the patterns are gradually smoothed. At \(a=1\), the system coincides with the ordinary FDNF-based FECA.

The rate at which this smoothing occurs is strongly rule-dependent. To quantify this effect, we use two simple scalar indicators. The first is the spatial contrast at time \(T\):
\begin{equation}
R_k(a;T)
= \max_{0\le j<N}x_j^T - \min_{0\le j<N}x_j^T.
\label{eq:contrast}
\end{equation}
The second is a fuzziness index:
\begin{equation}
B_k(a;T) =
\frac1N\sum_{j=0}^{N-1}4x_j^T(1-x_j^T).
\label{eq:fuzziness_index}
\end{equation}
The quantity \(B_k(a;T)\) is zero for binary configurations and becomes large when many sites take intermediate fuzzy values. Thus \(R_k\) measures spatial contrast, whereas \(B_k\) measures deviation from binary states.

\begin{figure}[hbt]
\centering
\includegraphics[width=0.8\textwidth]{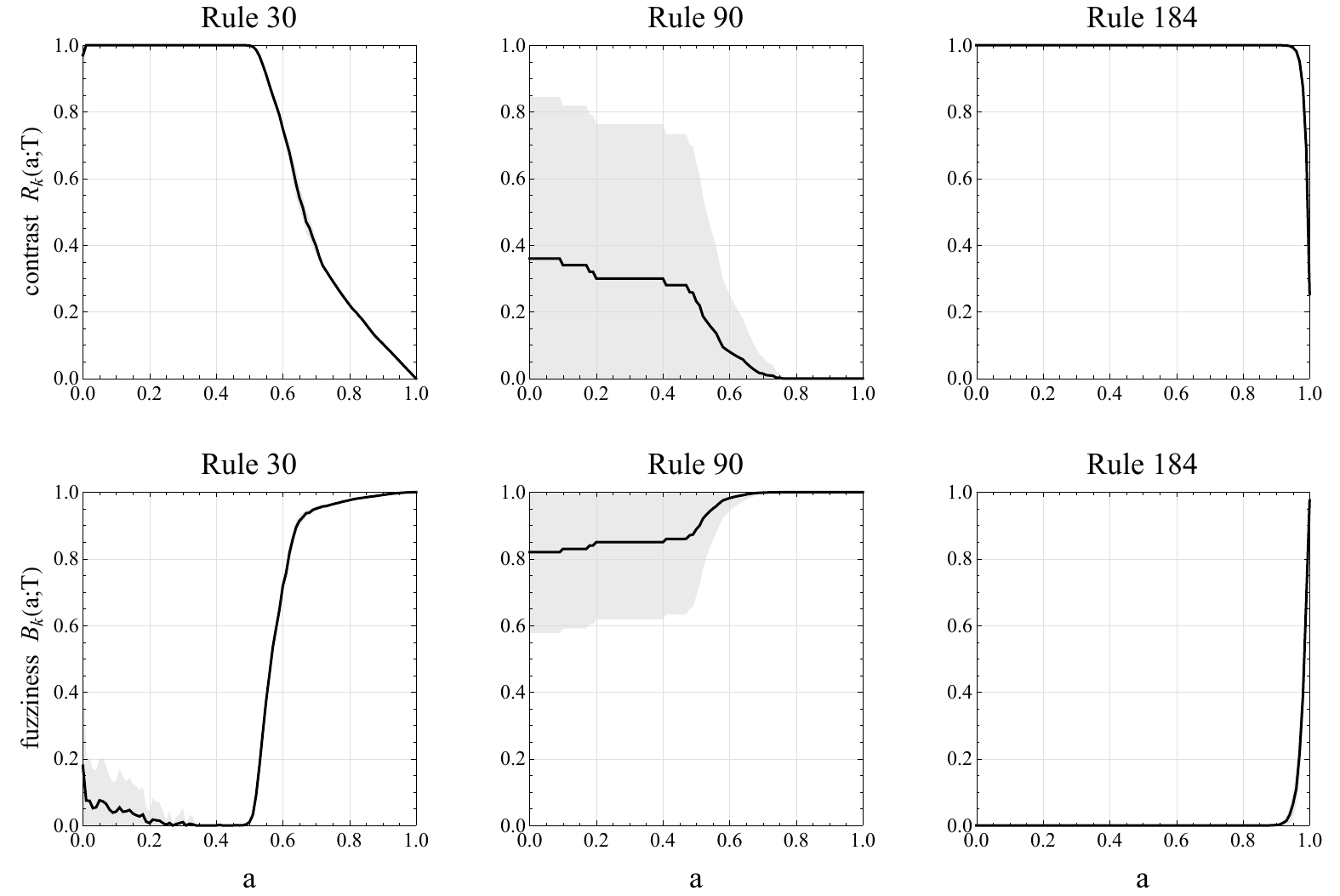}
\caption{
Quantitative characterization of pattern smoothing under the one-parameter fuzzification function \(q_a\). The simulations were performed on a periodic lattice of \(N=100\) sites up to \(T=100\), using 50 random initial conditions with values chosen independently from \([0,1]\). The parameter \(a\) was varied from 0 to 1 with step size 0.01. For each rule, the same ensemble of initial conditions was reused for all
parameter values. The upper panels show the contrast \(R_k(a;T)=\max_j x_j^T-\min_j x_j^T\), and the lower panels show the fuzziness index \(B_k(a;T)=N^{-1}\sum_j4x_j^T(1-x_j^T)\). Solid curves denote ensemble means, and gray bands indicate one standard deviation.
}
\label{fig:qa_contrast_fuzziness}
\end{figure}

Figure~\ref{fig:qa_contrast_fuzziness} shows these quantities as functions of \(a\). The results support the interpretation of ordinary FDNF as one endpoint of a broader rule-preserving family. Moving away from the FDNF endpoint can recover high-contrast ECA-like patterns, while moving toward the identity map produces FDNF-like smoothing.

%\FloatBarrier

%%%%%%%%%%%%%%%%%%%%%%%%%%%%%%%%%%%%%%
\subsection{Local explanation by boundary sensitivity}

The rule-dependent smoothing observed above can be partially explained by a linearization near binary ECA orbits.
Let
\(
s^t_j\in\{0,1\}
\)
be an orbit of the rule \(k\) ECA, and consider a small admissible perturbation
\(
x^t=s^t+\eta^t\in[0,1]^N.
\)
Since the one-sided endpoint derivatives of \(q_a\) satisfy \(q_a'(0)=q_a'(1)=a\), the perturbation dynamics near the binary orbit is controlled by the boundary slope \(a\) and the Boolean sensitivity of the underlying ECA rule.

\begin{proposition}%[Linearization near a binary ECA orbit]
Let \(s^t\) be a binary orbit of the rule \(k\) ECA, and let
\(x^t=s^t+\eta^t\in[0,1]^N\) be a sufficiently small admissible perturbation.
Then the GFECA defined by \(q_a\) satisfies
\[
\eta^{t+1}_j
=
a\left( \sigma^t_{j,L}\eta^t_{j-1} + \sigma^t_{j,C}\eta^t_j + \sigma^t_{j,R}\eta^t_{j+1} \right) + O(\|\eta^t\|_\infty^2),
\]
where
\begin{align*}
\sigma^t_{j,L}
&=
F_k(1,s^t_j,s^t_{j+1})-F_k(0,s^t_j,s^t_{j+1}),
\\
\sigma^t_{j,C}
&=
F_k(s^t_{j-1},1,s^t_{j+1})-F_k(s^t_{j-1},0,s^t_{j+1}),
\\
\sigma^t_{j,R}
&=
F_k(s^t_{j-1},s^t_j,1)-F_k(s^t_{j-1},s^t_j,0).
\end{align*}
In particular, each coefficient belongs to \(\{-1,0,1\}\).
\end{proposition}

\begin{proof}
The FDNF polynomial \(f_k\) is multi-affine. Therefore, at a vertex
\((\epsilon_1,\epsilon_2,\epsilon_3)\in\{0,1\}^3\), its partial derivative with respect to the first variable is
\[
\partial_x f_k(\epsilon_1,\epsilon_2,\epsilon_3)
=
F_k(1,\epsilon_2,\epsilon_3)-F_k(0,\epsilon_2,\epsilon_3),
\]
and analogous formulas hold for the other variables. Since the endpoint derivatives of \(q_a\) are understood as one-sided derivatives inside \([0,1]\) and satisfy \(q_a'(0)=q_a'(1)=a\), the chain rule gives the stated linearization.
\end{proof}

This proposition gives a local mechanism by which a affects the persistence or loss of binary-like patterns.
If the product of the Boolean sensitivity matrices along the ECA orbit has a large growth rate, fuzzy perturbations can be amplified unless \(a\) is sufficiently small. Thus the transition from ECA-like behavior to FDNF-like smoothing depends both on the fuzzification function and on the Boolean rule.
This type of linearized sensitivity is related to damage-spreading and Lyapunov-exponent approaches for cellular automata\cite{Bagnoli1992}.

%%%%%%%%%%%%%%%%%%%%%%%%%%%%%%%%%%%%%%%%%
%%%%%%%%%%%%
\subsection{Gap-induced pattern formation}
\label{subsec:gap_patterns}

The previous examples used continuous fuzzification functions. We now show that
discontinuous endpoint-preserving functions can also generate nontrivial patterns.
For \(1\le a\le 2\), define
\begin{equation}
r_a(x)
=
\begin{cases}
a x, & 0\le x\le \frac12,\\
a x-a+1, & \frac12 < x\le 1.
\end{cases}
\label{eq:gap_function}
\end{equation}
Then \(r_a\in\Omega\) for all \(1\le a\le 2\). When \(a=1\), \(r_a\) is the
identity map and the GFECA reduces to the ordinary FDNF. When \(a>1\), the
map has a downward jump at \(x=1/2\). The jump size is
\( \ 
r_a\left(\frac12-0\right)-r_a\left(\frac12+0\right)
=
a-1.
\)

In this experiment, we set
\( \ 
g=u=v=w=r_a.
\)
Although \(r_a\) is discontinuous for \(a>1\), the endpoint-preserving property is unchanged. Hence the resulting map remains a GFECA associated with the original ECA rule.

\begin{figure}[htb]
\centering
\includegraphics[width=0.95\textwidth]{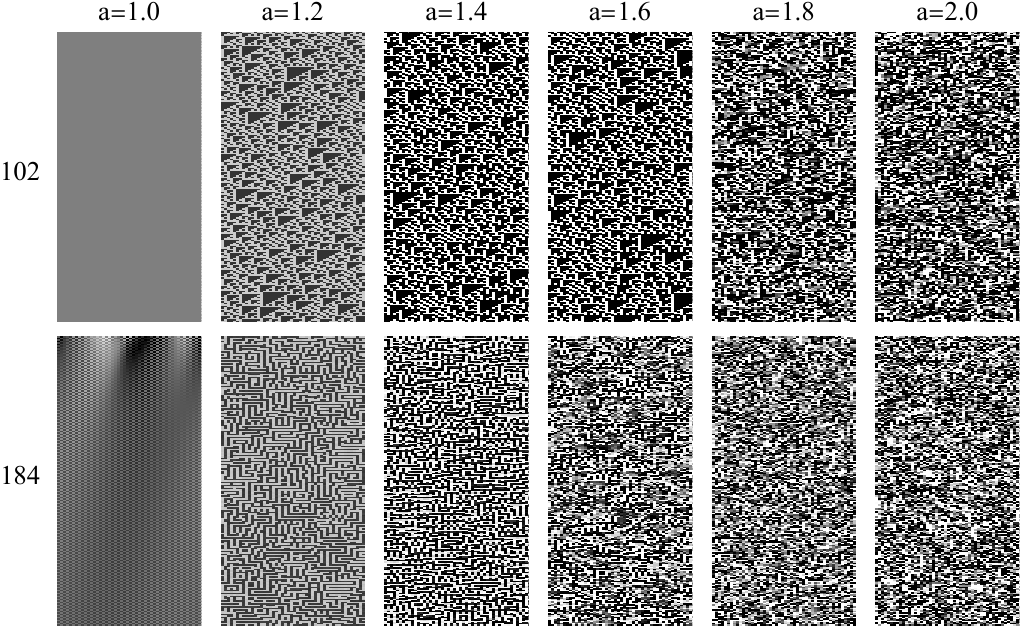}
\caption{
Time-evolution patterns of the GFECAs generated by the gap fuzzification function \(r_a\) in Eq.~\eqref{eq:gap_function}. The upper row shows rule 102 and the lower row shows rule 184. The simulations were performed on a periodic lattice of \(N=50\) sites for \(0\le t\le 200\). The same random initial condition in \([0,1]^N\) was used for all parameter values. The columns correspond to \(a=1.0,1.2,1.4,1.6,1.8,2.0\). The case \(a=1\) coincides with ordinary FDNF, whereas \(a>1\) introduces a discontinuity gap at \(x=1/2\). As the gap size increases, new pattern regimes emerge.
}
\label{fig:gap_patterns}
\end{figure}

Figure~\ref{fig:gap_patterns} shows the time-evolution patterns for rules 102 and 184. For \(a=1\), the systems reproduce the ordinary FDNF behavior. As the gap size increases, new spatial structures appear. In rule 102, ECA-like structures emerge for intermediate values of \(a\), whereas larger values of \(a\) produce more irregular patterns. In rule 184, an intermediate gap creates labyrinthine patterns that are not observed in the ordinary FDNF dynamics.

\begin{figure}[htb]
\centering
\includegraphics[width=0.7\textwidth]{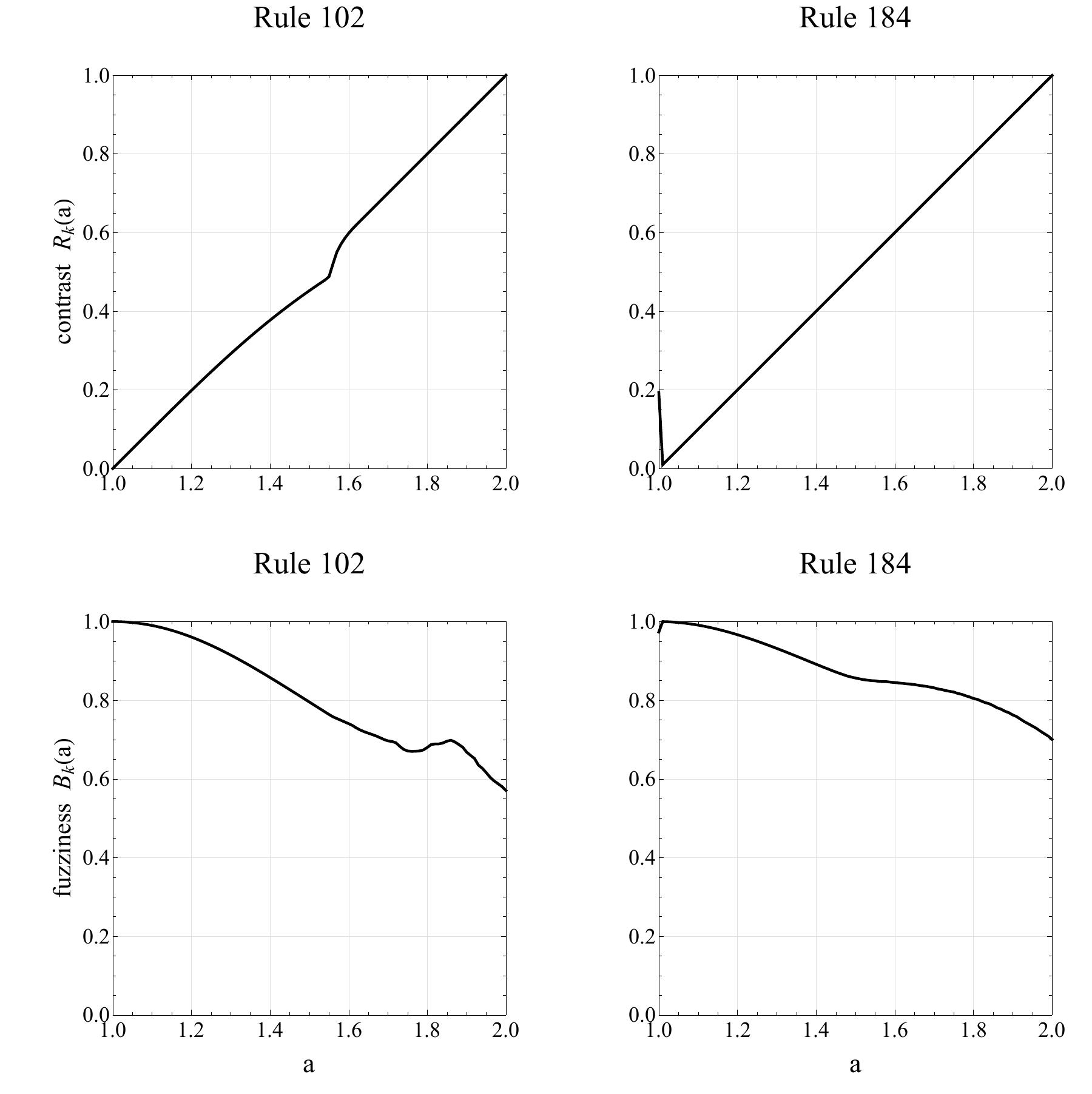}
\caption{
Quantitative characterization of the gap-induced patterns generated by
the fuzzification function \(r_a\). The simulations were performed on a periodic lattice of \(N=50\) sites up to \(T=200\), using 30 random initial conditions with values chosen independently from \([0,1]\). The parameter \(a\) was varied from 1 to 2 with step size 0.01. The upper panels show the contrast
\(
R_k(a)=\max_{(t,j)\in D}x_j^t-\min_{(t,j)\in D}x_j^t,
\)
and the lower panels show the fuzziness index
\(
B_k(a)=|D|^{-1}\sum_{(t,j)\in D}4x_j^t(1-x_j^t),
\)
where \(D=\{(t,j)\mid 100\le t\le 200,\ 0\le j<N\}\). 
Solid curves denote ensemble means. The standard deviations were smaller than the line width and are therefore not shown.
}
\label{fig:gap_metrics_RB}
\end{figure}

To quantify these gap-induced changes, we compute contrast and fuzziness on the late-time spacetime window
\[
D=\{(t,j)\mid 100\le t\le 200,\ 0\le j<N\}.
\]
For this window, define
\[
R_k(a)
=
\max_{(t,j)\in D}x_j^t
-
\min_{(t,j)\in D}x_j^t,
\]
and
\[
B_k(a)
=
\frac{1}{|D|}
\sum_{(t,j)\in D}4x_j^t(1-x_j^t).
\]
Figure~\ref{fig:gap_metrics_RB} shows the ensemble means over
30 random initial conditions. These quantities confirm that the discontinuity gap changes both the contrast of the pattern and the degree to which intermediate fuzzy values remain.

These examples show that discontinuities in fuzzification functions can act as branch-separating mechanisms. They do not merely recover the original ECA pattern; rather, they create new pattern regimes inside the generalized FDNF framework.

The examples in this section demonstrate that FDNF should be regarded as one distinguished endpoint inside a larger rule-preserving family. 
The ordinary FDNF corresponds to the identity fuzzification function and gives the canonical multi-affine extension. By replacing the identity with threshold-like, deformed,
or discontinuous endpoint-preserving functions, one can promote ECA-like recovery, tune the degree of smoothing, or generate new pattern regimes.
Thus, generalized FDNF provides a systematic way to study how fuzzy pattern dynamics depend on the shape and branch structure of the fuzzification function.

\FloatBarrier

%%%%%%%%%%%%%%%%%%%%%%%%%%%%%%%%%%%%%%%%%%%%%%%%%%
%  section 5
%%%%%%%%%%%%%%%%%%%%%%%%%%%%%%%%%%%%%%%%%%%%%%%%%%

\section{Quantitative characterization of pattern changes}
\label{sec:support_exponent}

The preceding section showed that the shape of the endpoint-preserving transformation strongly affects the resulting spatiotemporal patterns. 
The contrast \(R\) and the fuzziness index \(B\) quantify, respectively, the range of cell values and the degree to which intermediate fuzzy values remain.
These quantities, however, do not describe how the non-uniform amplitudes are distributed over a spacetime observation window.

In this section we use a normalized high-order moment indicator to summarize this amplitude distribution. The indicator is used only as a finite-resolution participation-type measure for comparing patterns computed under the same numerical protocol. 
It is interpreted as an effective support exponent, or equivalently as a dimension-scaled normalized participation index, rather than as a geometric or fractal dimension.

\subsection{Effective support exponent from normalized high-order moments}
\label{subsec:support_exponent_definition}

Let \(D\) be a finite spacetime analysis domain,
\[
D=\{(t,j)\mid t_0\le t\le t_1,\ 0\le j<N\},
\]
and let \(M=|D|\). For a parameter value \(\theta\), let \(\phi_i^{(\theta)}\ge 0\), \(i=1,\ldots,M\), denote a nonnegative amplitude field on \(D\). Here the index \(i\) represents a spacetime site \((t,j)\in D\). In the simplest case one may take \(\phi_i=x_j^t\). 
When the state contains a nearly uniform background, we instead use the deviation field
\[
\phi_{t,j}^{(\theta)}
=
\left|x_j^t-\overline{x}_D^{(\theta)}\right|,
\qquad
\overline{x}_D^{(\theta)}
=
\frac1M\sum_{(t,j)\in D}x_j^t .
\]
This removes the uniform component and measures the support of the non-uniform part of the pattern.

We define normalized weights $w_i^{(\theta)}$ by
\[
w_i^{(\theta)}
=
\frac{\left(\phi_i^{(\theta)}\right)^2}{S_2^{(\theta)}}, \quad
S_2^{(\theta)}
=
\sum_{i=1}^M\left(\phi_i^{(\theta)}\right)^2,
\quad
\sum_{i=1}^M w_i^{(\theta)}=1.
\]
For \(p>1\), the normalized participation index is
\[
H_p^{(\theta)}
=
-\frac{1}{p-1}
\log_M
\left(
\sum_{i=1}^M \left(w_i^{(\theta)}\right)^p
\right),
\qquad
0\le H_p^{(\theta)}\le 1.
\]
Since the observation domain \(D\) is a two-dimensional spacetime window, we display the dimension-scaled version, which we call a support exponent, 
\begin{equation}
s_\theta^{(p)}
=
2H_p^{(\theta)}
=
-\frac{2}{p-1}
\log_M
\left(
\sum_{i=1}^M \left(w_i^{(\theta)}\right)^p
\right).
\end{equation}
The factor \(2\) is only a normalization convention for a two-dimensional spacetime observation window: 
it makes a fully spread amplitude distribution have value \(2\), while an amplitude distribution supported on \(O(\sqrt M)\) sites has value close to \(1\). 
The normalized index \(H_p\) is a Rényi-type participation quantity \cite{Renyi1961}. 
It is related in form to generalized dimensions used in multifractal analysis \cite{Halsey1986}, but here no box-size scaling is performed and no geometric dimension is claimed.
%This scaling should not be interpreted as a claim that \(s_\theta^{(p)}\) is a geometric dimension.

If \(S_2^{(\theta)}=0\), we set \(s_\theta^{(p)}=0\) by convention. More generally, when the non-uniform amplitude \(S_2^{(\theta)}\) is extremely small, the support exponent should be interpreted with care, because the weights \(w_i^{(\theta)}\) are invariant under multiplication of all amplitudes by a common constant. 
In the numerical interpretation below, we therefore read \(s_\theta^{(p)}\) together with the contrast \(R\), the fuzziness index \(B\), and the amplitude normalization \(S_2^{(\theta)}\). 
The support exponent is used only as a finite-resolution indicator of how broadly the non-uniform amplitude is distributed.

For equal nonzero amplitudes supported on exactly \(K\) sites, the above definition gives
\[
s_\theta^{(p)}=2\log_M K.
\]
Consequently, a single active site gives \(s=0\), a support of size \(K=\sqrt M\) gives \(s=1\), and a support of size \(K=M\) gives \(s=2\).
However, a connected line of \(\sqrt M\) active sites and \(\sqrt M\) active sites randomly scattered over \(D\) have the same value. This synthetic benchmark shows both the usefulness and the limitation of the measure:
it quantifies amplitude support, not geometric organization.

The numerical settings used for the main-text support-exponent calculations are summarized in Table~\ref{tab:support_settings}. In all cases, periodic boundary conditions are imposed. The same ensemble of initial conditions was reused for all parameter values within each ensemble experiment. 
Values smaller than \(10^{-12}\) and values larger than \(1-10^{-12}\) were rounded to \(0\) and \(1\), respectively. In the main-text calculations below, the moment order is fixed at \(p^*=16\). 
Therefore, the absolute values of \(s^{(p^*)}\) should be compared only within the same numerical protocol.

\begin{table}[!htb]
\centering
\small
\caption{Numerical settings used for the support-exponent calculations in the main text.}
\label{tab:support_settings}
%\label{tab:quantitative_parameters}
\begin{tabular}
%{p{0.22\textwidth}p{0.26\textwidth}p{0.27\textwidth}p{0.20\textwidth}}
{p{0.20\textwidth}p{0.24\textwidth}p{0.25\textwidth}p{0.18\textwidth}}
\hline
Experiment & Rules and parameter grid & System size and analysis domain
& Pattern field and moment order \\
\hline
\(q_a\) deformation
&
\(k=30,90,184\);
\(0\le a\le1\), \(\Delta a=0.01\)
&
\(N=100\), \(T=100\);
\(D:\ 50\le t\le100,\ 0\le j<N\)
&
\(\phi_{t,j}=|x_j^t-\overline{x}_D|\);
\(p^*=16\)
\\[1mm]
Gap function \(r_a\)
&
\(k=102,184\);
\(1\le a\le2\), \(\Delta a=0.01\)
&
\(N=50\), \(T=200\);
\(D:\ 100\le t\le200,\ 0\le j<N\)
&
\(\phi_{t,j}=|x_j^t-\overline{x}_D|\);
\(p^*=16\)
\\
\hline
\end{tabular}

\vspace*{3mm}
\begin{minipage}{0.95\textwidth}
For the \(q_a\)-deformation experiments, 50 independent random initial conditions with i.i.d. values in \([0,1]\) were used. For the gap-function experiments, 30 independent random initial conditions with i.i.d. values in \([0,1]\) were used.
\end{minipage}
\end{table}

%%%%%%%%%%%%%%%%%%%%%%%%%%%%%%%%%%%
%%%%%%%%%%%%%%

\subsection{Support exponent for the deformation from ECA-like behavior to FDNF}
\label{subsec:effective_dimension_qa}

We first apply the support exponent to the one-parameter family \(q_a\) introduced in Eq.~\eqref{eq:qa}. This family connects the threshold-like transformation at \(a=0\) to the identity map at \(a=1\), and hence interpolates between ECA-like behavior and ordinary FDNF.

For each rule \(k\in\{30,90,184\}\) and each parameter value \(a\in[0,1]\), we compute the spacetime pattern \(x_j^t(a)\) generated by the corresponding GFECA. Since ordinary FDNF may produce a nearly uniform background, we use the deviation field
\[
\phi_{t,j}^{(a)}
=
\left|x_j^t(a)-\overline{x}_D^{(a)}\right|.
\]
We then compute
\[
s_k^{\mathrm{dev}}(a)=s_a^{(p^*)},
\qquad p^*=16,
\]
from the definition above.

\begin{figure}[htb]
\centering
\includegraphics[width=\textwidth]{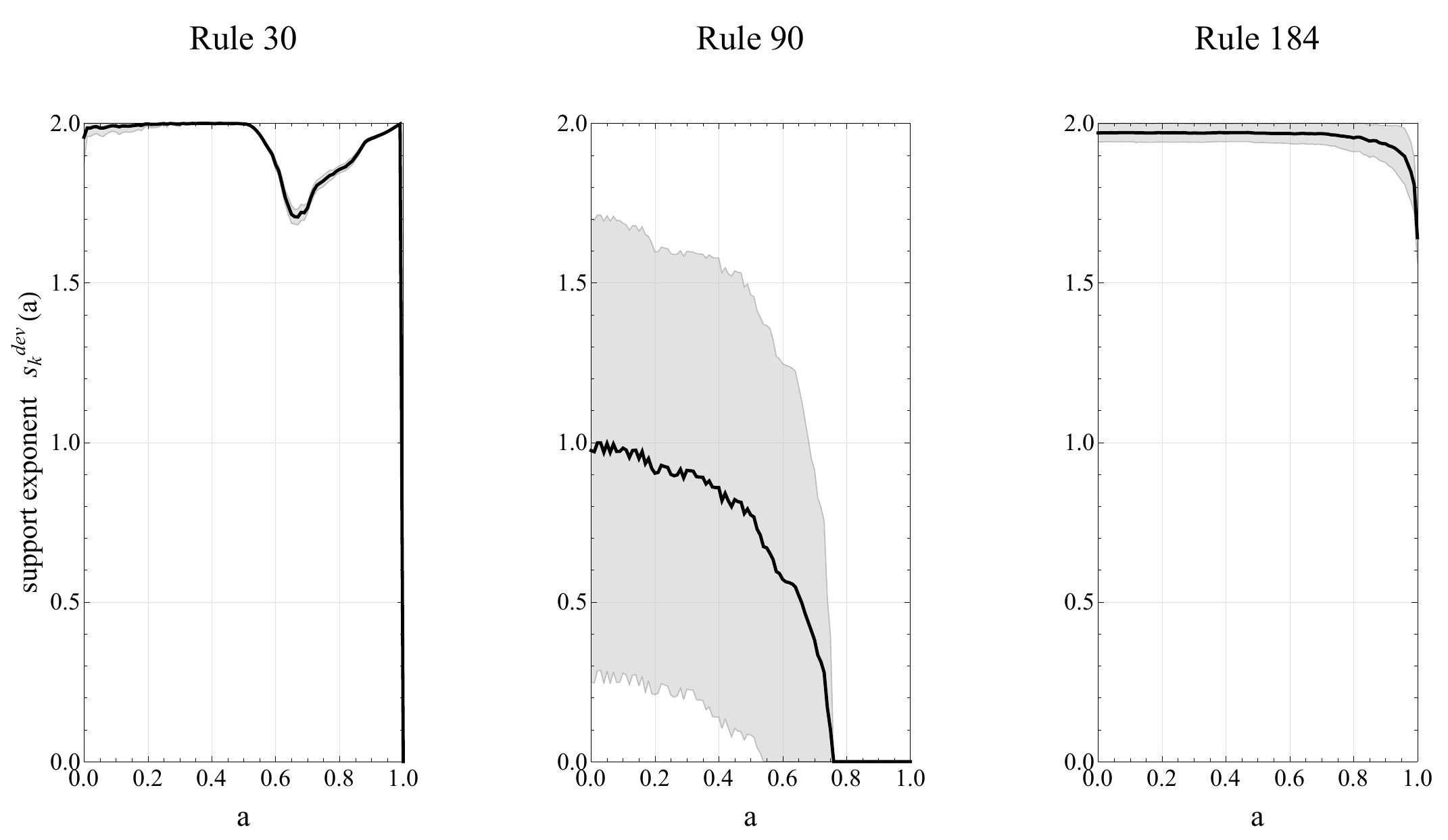}
\caption{The displayed quantity is the dimension-scaled participation index \(s=2H_p\) with \(p^*=16\). Solid curves denote ensemble means, and gray bands indicate one standard deviation.}
\label{fig:support_qa}
\end{figure}

Figure~\ref{fig:support_qa} shows the resulting support exponents for rules 30, 90, and 184. The curves should be interpreted together with the contrast and fuzziness indices in Fig.~\ref{fig:qa_contrast_fuzziness}.
Contrast and fuzziness describe the value distribution, whereas \(s_k^{\mathrm{dev}}(a)\) describes how broadly the non-uniform amplitude is spread over the observation window.

For small \(a\), the threshold-like transformation tends to preserve ECA-like structures. As \(a\) approaches \(1\), the system moves toward the ordinary FDNF endpoint.
The dip for rule 30 near \(a\simeq0.6\) reflects an intermediate smoothing
regime in which the non-uniform deviation field is relatively concentrated.
Since \(s^{(p)}\) is normalized by \(S_2\), the subsequent increase near the
FDNF endpoint should be interpreted together with the decreasing amplitude
normalization \(S_2/M\). 
The support exponent provides a finite-resolution way to describe how the amplitude support of the residual non-uniform part changes during this deformation. The results also show that the response is rule-dependent.

%%%%%%%%%%%%%%%%%%%%%%%%%%%%%%%%%%%%%%%%%%%%%%%

\subsection{Support exponent for gap-induced patterns}
\label{subsec:effective_dimension_gap}

We next consider the discontinuous gap transformation \(r_a\) defined in Eq.~\eqref{eq:gap_function}. The parameter \(a\in[1,2]\) controls the size of the jump discontinuity at \(x=1/2\). As shown in Fig.~\ref{fig:gap_patterns}, increasing the gap size can create characteristic spatiotemporal structures that are not observed in the ordinary FDNF case.

For this experiment, we compute the support exponent for rules 102 and 184. The analysis domain \(D\) is a late-time spacetime window, and we again use the deviation field
\[
\phi_{t,j}^{(a)}
=
\left|x_j^t(a)-\overline{x}_D^{(a)}\right|.
\]
The corresponding exponent is denoted by
\[
s_k^{\mathrm{gap}}(a), \qquad k\in\{102,184\},
\]
with \(p^*=16\).

\begin{figure}[htb]
\centering
\includegraphics[width=0.45\textwidth]{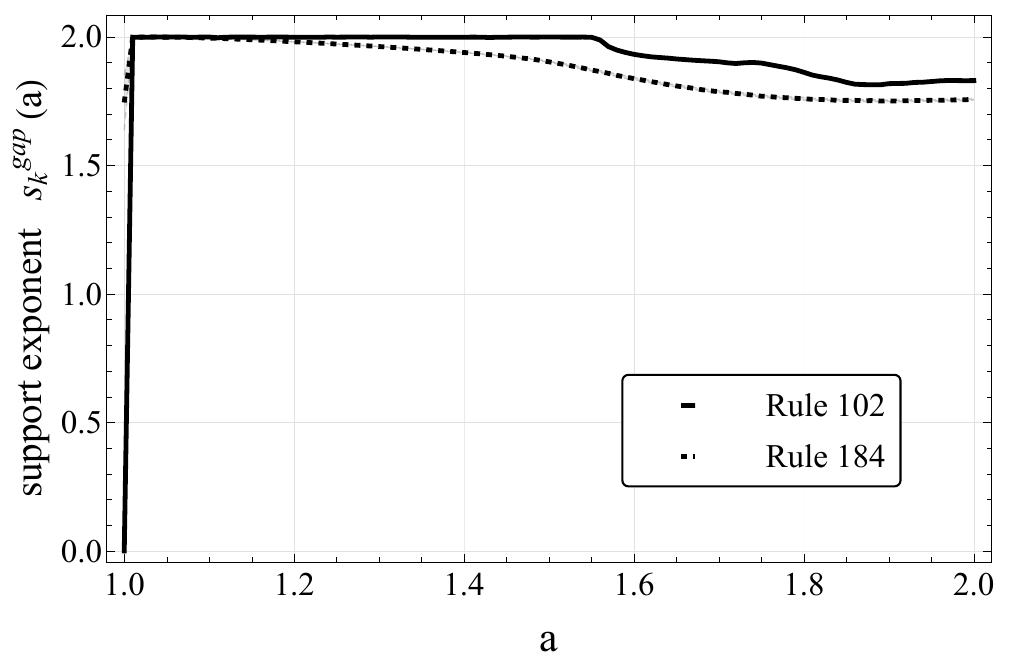}
\caption{Effective support exponent of the gap-induced patterns generated by the transformation \(r_a\) in Eq.~\eqref{eq:gap_function}. The displayed quantity is \(s=2H_p\) with \(p^*=16\). The deviation field is used to remove uniform backgrounds. The numerical settings are given in Table~\ref{tab:support_settings}.}
\label{fig:support_gap}
\end{figure}

Figure~\ref{fig:support_gap} shows \(s_{102}^{\mathrm{gap}}(a)\) and \(s_{184}^{\mathrm{gap}}(a)\). Together with the contrast and fuzziness indices, this quantity gives a finite-resolution description of how the non-uniform amplitude support changes as the gap-induced patterns pass from nearly uniform states to more widely spread patterns. 
Since the deviation field is used, values close to the uniform FDNF endpoint should be interpreted together with the amplitude level of the non-uniform component.

%%%%%%%%%%%%%%%%%%%%%%%%%%%%%%%%%%%%%%%%%%%
%%%%%%%%%%%%%%%%%%%%%%%%%%%%%%%%%%%%%%%%%%%%%

\subsection{Summary of quantitative indicators}
\label{subsec:indicator_summary}

The contrast \(R\), the fuzziness index \(B\), and the support exponent \(s\) play complementary roles in describing the pattern changes generated by endpoint-preserving transformations. The contrast measures the range of cell values, the fuzziness index measures the degree to which intermediate fuzzy values remain, and the support exponent measures the effective amplitude support in a finite spacetime observation window.

For the \(q_a\)-deformation and the gap family \(r_a\), these indicators provide a compact finite-resolution description of the transition from ECA-like recovery to FDNF-like smoothing and of the emergence of gap-induced pattern regimes. A moment-order robustness check for \(p=8,16,32,64\), together with the corresponding amplitude normalizations \(S_2/M\), is provided in Appendix~\ref{app:moment_order_robustness}.

%%%%%%%%%%%%%%%%%%%%%%%%%%%%%%%%%%%%%%%%%%%%%%%%%
%  section 6
%%%%%%%%%%%%%%%%%%%%%%%%%%%%%%%%%%%%%%%%%%%%%%%%
\section{Minimal three-cell GFECAs and nonlinear dynamics}
\label{sec:minimal_dynamics}

We now turn to the minimal periodic lattice of size \(N=3\).
Although a Boolean three-cell ECA has only finitely many states, so every Boolean orbit is eventually periodic, the corresponding GFECA defines a map on the unit cube \([0,1]^3\). 
This minimal setting already exhibits nonlinear dynamics that cannot occur in the finite Boolean system.

Throughout this section we use the non-monotone expanding fuzzification function
\begin{equation}
u_*(x)=
\begin{cases}
3x, & 0\le x\le \dfrac13,\\[1mm]
-3x+2, & \dfrac13 < x\le \dfrac23,\\[1mm]
3x-2, & \dfrac23 < x\le 1,
\end{cases}
\label{eq:u_star}
\end{equation}
which is the function shown in Fig.~\ref{fig:fuzzification_functions}(f). 

Let \(\widetilde f_k\) be the GFDNF rule constructed from the FDNF polynomial \(f_k\) with
\( g=\mathrm{id},\qquad u=v=w=u_*\). 
On the periodic lattice of size \(3\), write the state as \((x,y,z)\in[0,1]^3\). 
The corresponding three-cell GFECA map is
\begin{equation}
T_k(x,y,z)
=
\left(
\widetilde f_k(z,x,y),
\widetilde f_k(x,y,z),
\widetilde f_k(y,z,x)
\right).
\label{eq:three_cell_map}
\end{equation}
The cyclic order reflects the radius-one neighborhood on the three-site periodic lattice.

We use representative rules to show two mechanisms by which expanding endpoint-preserving transformations create nonlinear dynamics. 
First, in rules such as 51 and 85, expanding one-dimensional dynamics is inherited directly from the fuzzification function. 
Second, in rule 210, the interaction between the ECA rule and \(u_*\) creates locally stable period-six cycles together with an expanding invariant line set. 
The coexistence of these objects leads to finely mixed finite-resolution class structures.

\begin{figure}[htb]
\centering
\includegraphics[width=0.55\textwidth]{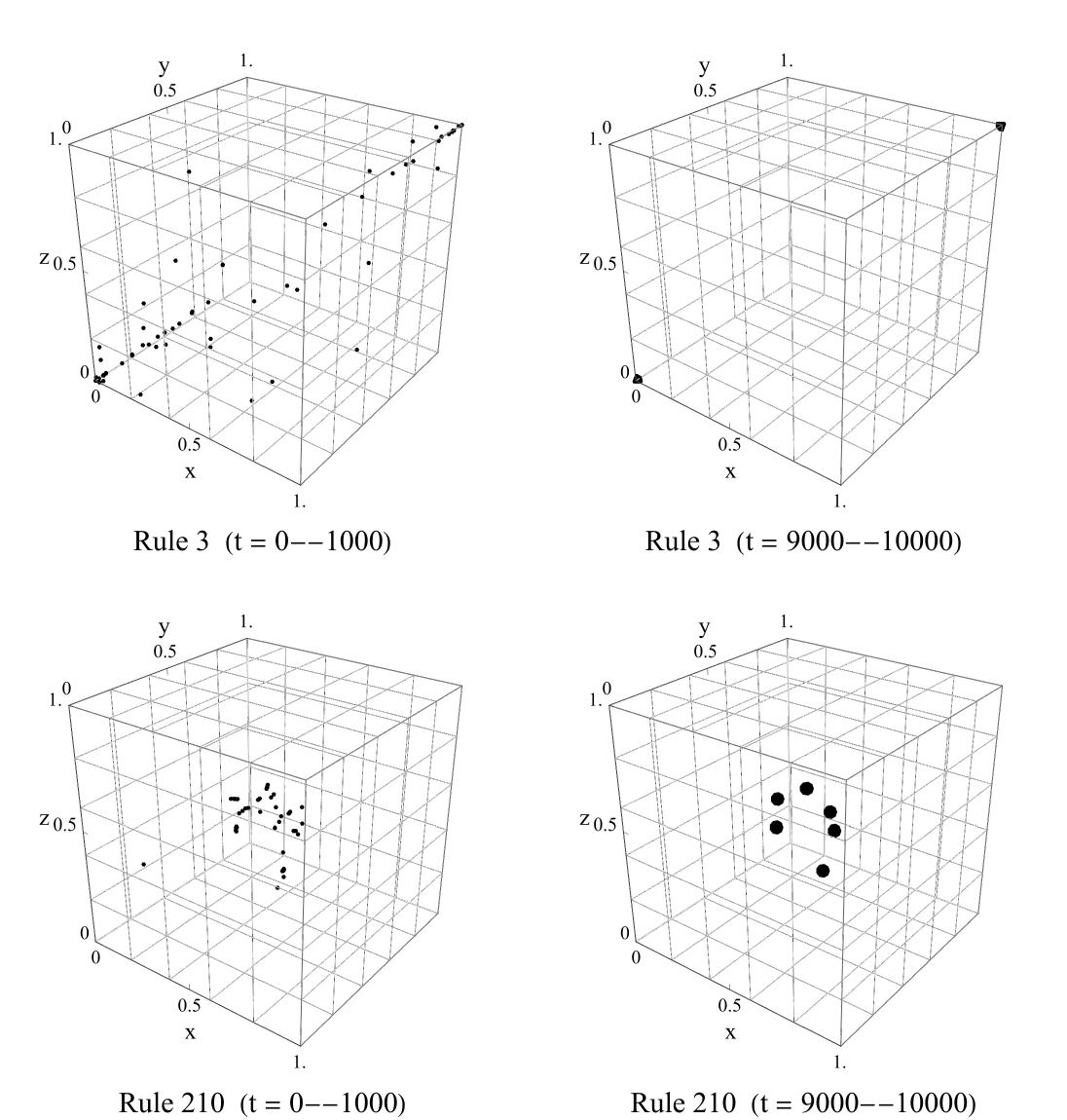}
\caption{
Representative three-cell GFECA orbits generated by the expanding fuzzification function \(u_*\). 
For the displayed non-binary initial condition, rule 3 approaches a period-two orbit, whereas rule 210 approaches one of the period-six cycles constructed below. 
The left panels show transient orbit samples for \(0\le t\le1000\), and the right panels show late-time samples for \(9000\le t\le10000\).
}
\label{fig:three_cell_orbits_rule3_rule210}
\end{figure}

Figure~\ref{fig:three_cell_orbits_rule3_rule210} shows representative three-cell orbit samples for rules 3 and 210. 
For the displayed non-binary initial condition, rule 3 approaches a simple period-two orbit, whereas rule 210 approaches one of the period-six cycles constructed below. 
These examples are intended only as representative orbit samples; they are not global classification results.

\FloatBarrier

%%%%%%%%%%%%%%%%%%%%%%%%%%%%%%%%%%%%%%%%%%%%%%%%%%%%%%%%%%%%%%%%%
\subsection{Rule 210: symmetries and an expanding invariant line set}
\label{subsec:rule210_symmetry}

We first analyze rule 210. Under the rule-number convention used in this paper, the FDNF polynomial of rule 210 is
\begin{equation}
f_{210}(X,Y,Z)=2XYZ-2XZ+X-YZ+Z.
\label{eq:rule210_f210}
\end{equation}
With the fuzzification function \(u_*\), the corresponding GFDNF local rule is
\[
\widetilde f_{210}(x,y,z)
=
f_{210}(u_*(x),u_*(y),u_*(z)).
\]
Let
\[
\rho(x,y,z)=(y,z,x)
\]
be the cyclic coordinate permutation. 
Then the three-cell map \(T=T_{210}\) satisfies
\begin{equation}
T\circ\rho=\rho\circ T.
\label{eq:rho_equivariance}
\end{equation}
We also use the complement-reversal map
\[
\kappa(x,y,z)=(1-z,1-y,1-x).
\]
The rule 210 map satisfies the shifted complement relation
\begin{equation}
T\circ\kappa=\rho\circ\kappa\circ T.
\label{eq:kappa_relation}
\end{equation}
These two symmetries organize the period-six cycles constructed below.

\begin{proposition}%[Symmetry of the rule 210 GFDNF rule]
\label{prop:rule210_symmetry}
The local rule \(\widetilde f_{210}\) satisfies
\begin{equation}
\widetilde f_{210}\left(\frac12,y,z\right)=\frac12
\label{eq:rule210_half_identity}
\end{equation}
for all \(y,z\in[0,1]\). It also satisfies
\begin{equation}
\widetilde f_{210}(x,y,z)
=
1-\widetilde f_{210}(1-x,1-z,1-y)
\label{eq:rule210_complement_identity}
\end{equation}
for all \(x,y,z\in[0,1]\).
\end{proposition}

\begin{proof}
First, \(u_*(1/2)=1/2\). Substituting \(X=1/2\) into
\eqref{eq:rule210_f210}, we obtain
\[
f_{210}\left(\frac12,Y,Z\right)
=
2\cdot\frac12YZ-2\cdot\frac12Z+\frac12-YZ+Z
=
\frac12.
\]
This proves \eqref{eq:rule210_half_identity}. Next,
\( \ 
u_*(1-x)=1-u_*(x)\), 
and a direct calculation using \eqref{eq:rule210_f210} gives
\(\ 
f_{210}(X,Y,Z)
=
1-f_{210}(1-X,1-Z,1-Y)\). 
Combining these identities gives \eqref{eq:rule210_complement_identity}.
\end{proof}

Equation~\eqref{eq:rule210_half_identity} implies that rule 210 contains a cyclic invariant line set
\begin{equation}
\begin{split}
\mathcal{L}
=
&\{(x,1/2,1/2)\mid 0\le x\le1\}
\cup
\{(1/2,x,1/2)\mid 0\le x\le1\}\\
&\qquad
\cup
\{(1/2,1/2,x)\mid 0\le x\le1\}.
\end{split}
\label{eq:line_set_L}
\end{equation}
These are the three line segments through \((1/2,1/2,1/2)\) parallel to the coordinate axes. Starting from a point of the form \((x_0,1/2,1/2)\), the orbit moves cyclically through these three line segments:
\[
(x_0,1/2,1/2)
\mapsto
(1/2,x_1,1/2)
\mapsto
(1/2,1/2,x_2)
\mapsto
(x_3,1/2,1/2)
\mapsto\cdots.
\]
The scalar dynamics along this cyclic invariant line set is
\begin{equation}
x_{n+1}=h(x_n):=\frac14+\frac12u_*(x_n).
\label{eq:h_rule210}
\end{equation}
Since \(|h'(x)|=3/2\) on each branch where \(h\) is differentiable, generic orbits in this invariant line set have Lyapunov exponent
\(\  \log 3/2 >0\). 
Thus rule 210 contains an expanding one-dimensional invariant mechanism.
At the same time, as shown below, the full three-dimensional map also has locally stable period-six cycles.

%%%%%%%%%%%%%%%%%%%%%%%%%%%%%%%%%%%%%%%%%%%%%%%%%%%%%%%%%%%%%%%%%
\subsection{Stable period-six cycles of rule 210}
\label{subsec:rule210_cycles}

Let \(T=T_{210}\) be the three-cell map defined in \eqref{eq:three_cell_map}. We now construct a period-six orbit of \(T\). 
Consider the six points
\[
\begin{aligned}
P_0&=(a^*,1/2,b^*),&
P_1&=(c^*,d^*,1/2),&
P_2&=(1/2,b^*,a^*),\\
P_3&=(d^*,1/2,c^*),&
P_4&=(b^*,a^*,1/2),&
P_5&=(1/2,c^*,d^*).
\end{aligned}
\]
The algebraic numbers \(a^*,b^*,c^*,d^*\) are specified and enclosed by rational intervals in \ref{app:rule210_interval}. For reference,
\[
a^*\simeq0.4785350944,
\qquad
b^*\simeq0.6235572234,
\]
\[
c^*\simeq0.2907948676,
\qquad
d^*\simeq0.5560666557.
\]
The interval verification in \ref{app:rule210_interval} gives the ordering
\begin{equation}
c^*<\frac13<a^*<\frac12<d^*<b^*<\frac23 .
\label{eq:abcd_ordering}
\end{equation}
Hence the relevant branches of \(u_*\) are
\[
u_*(c^*)=3c^*,\quad
u_*(a^*)=2-3a^*,
\quad
u_*(b^*)=2-3b^*,
\quad
u_*(d^*)=2-3d^*.
\]
In particular, the branch itinerary of the orbit is well defined and stays away from the break points \(1/3\) and \(2/3\) of \(u_*\).

\begin{proposition}%[A period-six orbit of rule 210]
\label{prop:rule210_period_six}
Let \(a^*,b^*,c^*,d^*\) be the algebraic numbers specified in \ref{app:rule210_interval}. 
Then
\[
T(P_i)=P_{i+1},\qquad i=0,\ldots,5,
\]
where indices are taken modulo \(6\). 
Moreover, the six points are mutually distinct. 
Hence they form a period-six orbit of the rule 210 three-cell GFECA.
\end{proposition}

\begin{proof}
The identities \(T(P_i)=P_{i+1}\) are verified in \ref{app:rule210_interval} by substituting the algebraic relations defining \(a^*,b^*,c^*,d^*\) on the branches specified by \eqref{eq:abcd_ordering}. The cyclic equivariance \eqref{eq:rho_equivariance} implies that it is enough to verify two successive transitions. 
The intervals in \eqref{eq:abcd_ordering} are
mutually disjoint, so the six points are distinct. Therefore the orbit has minimal period six.
\end{proof}

By the complement-reversal symmetry \eqref{eq:kappa_relation}, there is a second period-six orbit. Write \(\overline{x}=1-x\) and define
\[
\begin{aligned}
P'_0&=(\overline{b^*},1/2,\overline{a^*}),&
P'_1&=(\overline{d^*},\overline{c^*},1/2),&
P'_2&=(1/2,\overline{a^*},\overline{b^*}),\\
P'_3&=(\overline{c^*},1/2,\overline{d^*}),&
P'_4&=(\overline{a^*},\overline{b^*},1/2),&
P'_5&=(1/2,\overline{d^*},\overline{c^*}).
\end{aligned}
\]
Equivalently,
\(\ 
P'_i=\rho^i\kappa(P_i)
\). 
The relation \eqref{eq:kappa_relation} implies that
\(P'_0,\ldots,P'_5\) also form a period-six orbit.

\begin{proposition}%[Local asymptotic stability]
\label{prop:rule210_stability}
The period-six orbit \(P_0,\ldots,P_5\) of the rule 210 three-cell GFECA is locally asymptotically stable. The symmetric orbit \(P'_0,\ldots,P'_5\) has the same local stability.
\end{proposition}

\begin{proof}
Let
\[
M=DT(P_5)DT(P_4)\cdots DT(P_0)
\]
be the Jacobian matrix of \(T^6\) at \(P_0\). Since the interval bounds in \ref{app:rule210_interval} show that the orbit stays a positive distance away from the break points \(1/3\) and \(2/3\) of \(u_*\), the branch itinerary is locally constant near the orbit. 
Hence \(T^6\) is differentiable in a neighborhood of \(P_0\).

\ref{app:rule210_interval} encloses the algebraic coordinates of the period-six orbit by rational intervals and verifies, using interval arithmetic in the sense of Moore~\cite{Moore1966}, that the characteristic polynomial of \(M\) satisfies the Jury stability criterion~\cite{Jury1964}.
Therefore all eigenvalues of \(M\) lie strictly inside the unit circle. 
Hence the fixed point of \(T^6\) corresponding to \(P_0\) is locally asymptotically stable, and the period-six orbit of \(T\) is locally asymptotically stable. 
The symmetric orbit has the same stability by \eqref{eq:kappa_relation}.
\end{proof}

For reference, the eigenvalues of \(M\) are approximately
\[
\lambda_{1,2}\simeq0.37235662\pm0.56140204\,i,
\qquad
\lambda_3\simeq0.00284215.
\]

%%%%%%%%%%%%%%%%%%%%%%%%%%%%%%%%%%%%%%%%%%%%%%%%%%%%%%%%%%%%%%%%%
\subsection{Three-class finite-resolution slice classification}
\label{subsec:rule210_threeclass}

The rule 210 map contains two locally stable period-six cycles and the cyclic invariant line set \(\mathcal L\). 
Therefore, the finite-resolution slice structure should not be described only as a two-basin structure. 
We use the following three labels:
\[
B_+:\quad \text{convergence to } P_0,\ldots,P_5,
\]
\[
B_-:\quad \text{convergence to } P'_0,\ldots,P'_5,
\]
and
\[
\mathcal L:\quad \text{finite-resolution proximity to the cyclic invariant line set}.
\]
A residual label ``other'' is retained for diagnostic purposes, but no sampled point in the computations below was assigned to this residual class.

For a point \(X=(x,y,z)\), define the distance to the line set \(\mathcal L\) by
\begin{equation}
\begin{split}
&d_{\mathcal L}(X)
= \min\left\{
\max\left(\left|y-\frac{1}{2}\right|,\left|z-\frac{1}{2}\right|\right),\right. \\
&\left. \qquad \qquad \qquad
\max\left(\left|x-\frac{1}{2}\right|,\left|z-\frac{1}{2}\right|\right),
\max\left(\left|x-\frac{1}{2}\right|,\left|y-\frac{1}{2}\right|\right)
\right \}.  
\end{split}
\label{eq:line_set_distance}
\end{equation}
The numerical classification was performed as follows. A point was classified as \(B_+\) or \(B_-\) if its orbit satisfied the finite-time convergence criterion to the corresponding period-six cycle. 
Points not classified into either cycle were then tested for finite-resolution proximity to \(\mathcal L\) using \(d_{\mathcal L}\). 
Details of the numerical protocol, including tolerances, maximum iteration time, and sampling procedure, are given in Appendix~\ref{app:numerical_protocols}.

Figures~\ref{fig:rule210_threeclass_z05} and \ref{fig:rule210_threeclass_z03} show the three-class classification on the slices \(z=1/2\) and \(z=0.3\), respectively. 
On the slice \(z=1/2\), the line-set class occupies approximately one half of the sampled points, while the two stable period-six cycle classes occupy approximately one quarter each. 
On the slice \(z=0.3\), the three classes are nearly balanced. 
In both slices, no residual ``other'' class was observed under the numerical classification criterion.

\begin{figure}[htb]
\centering
\includegraphics[width=0.80\textwidth]{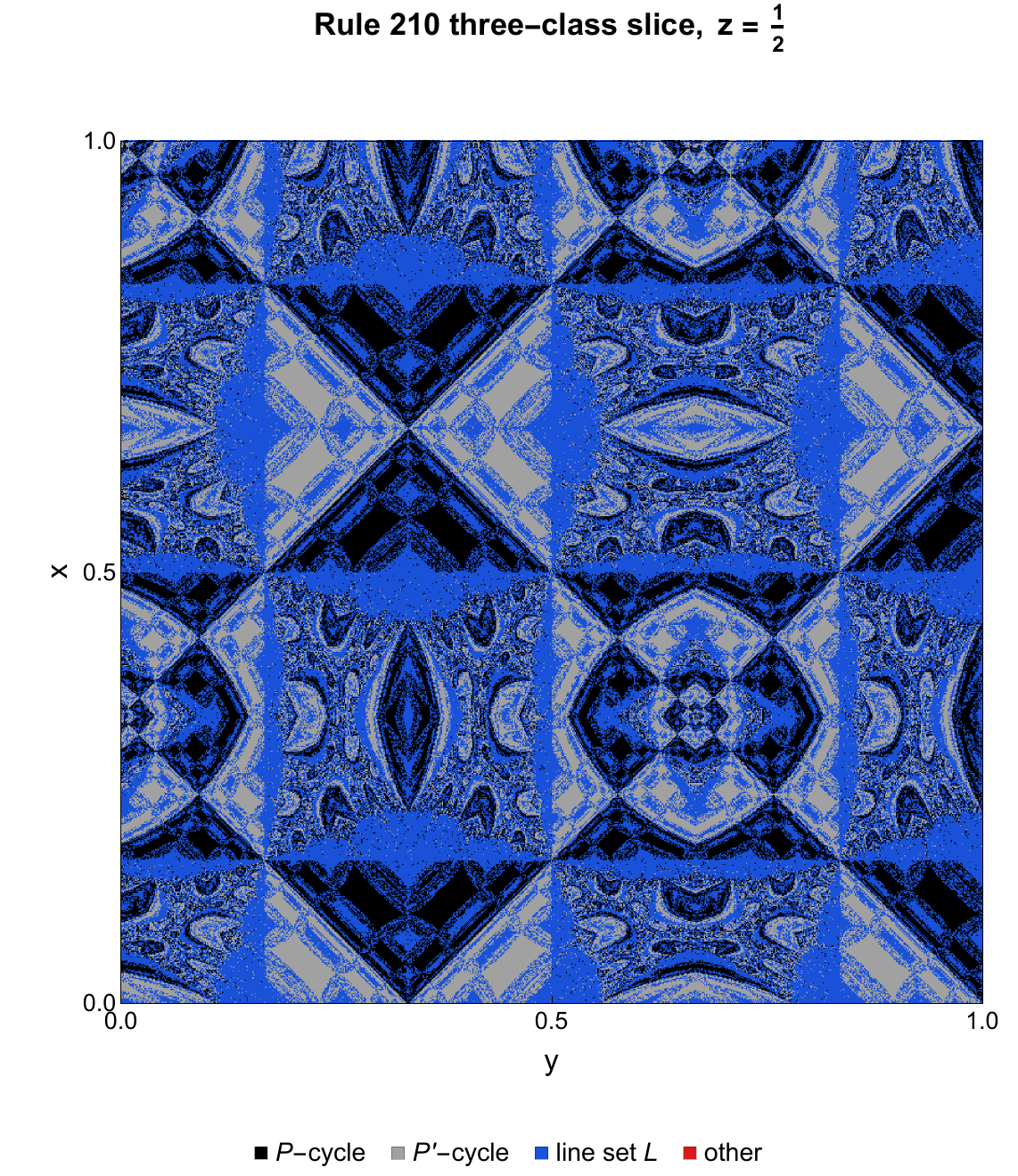}
\caption{
Three-class finite-resolution classification of the rule 210 three-cell GFECA on the slice \(z=1/2\). 
Black denotes convergence to the period-six cycle \(B_+\), gray denotes convergence to the symmetric period-six cycle \(B_-\), blue denotes finite-resolution proximity to the cyclic invariant line set \(\mathcal L\), and red would denote the residual ``other'' class. 
No sampled point was assigned to the residual class. 
The sampled fractions were approximately \(B_+\simeq0.250\), \(B_-\simeq0.250\), and \(\mathcal L\simeq0.500\).
}
\label{fig:rule210_threeclass_z05}
\end{figure}

\begin{figure}[htb]
\centering
\includegraphics[width=0.80\textwidth]{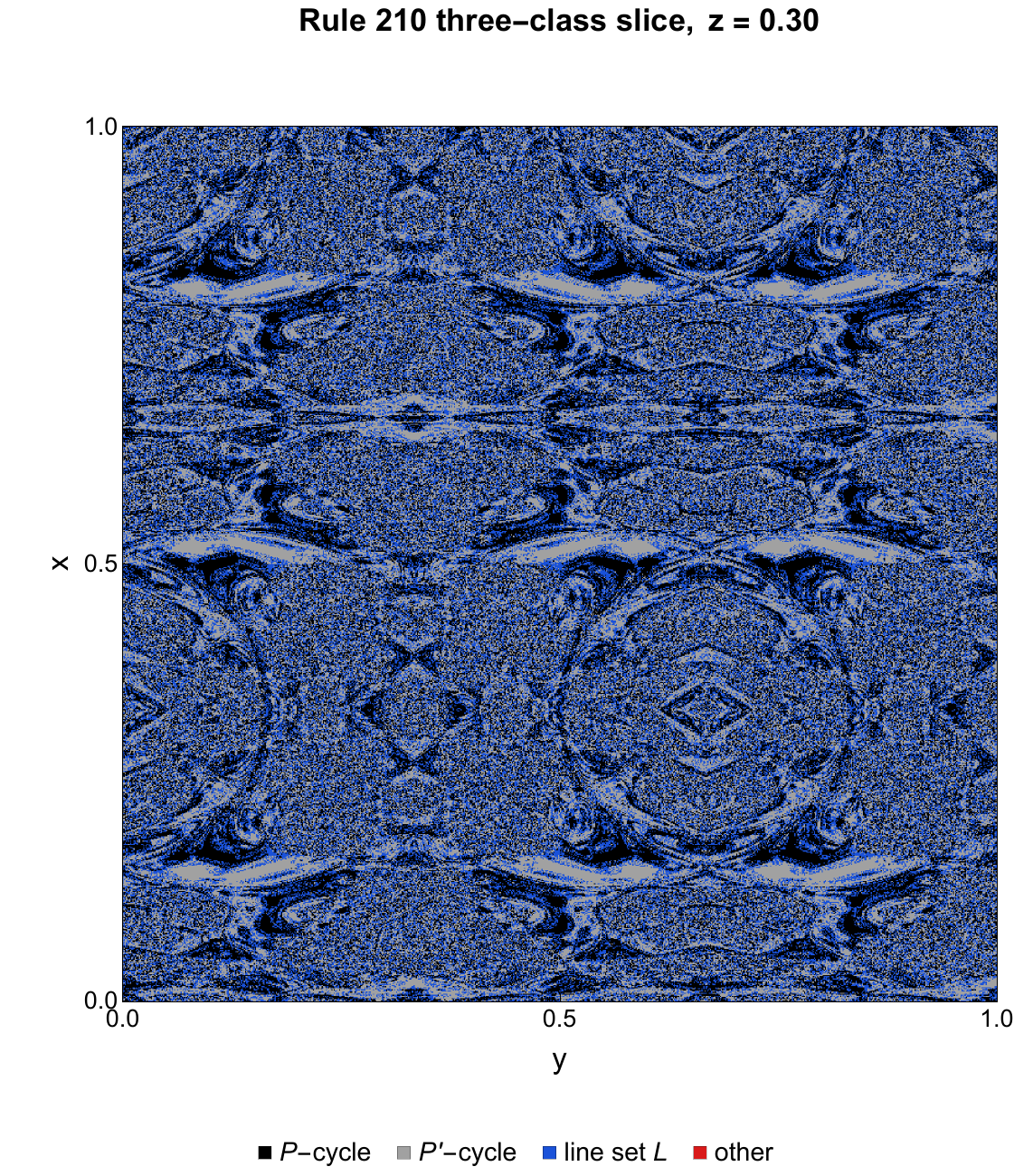}
\caption{
Three-class finite-resolution classification on the slice \(z=0.3\).
The same color convention as in Fig.~\ref{fig:rule210_threeclass_z05} is used.
The three classes are more finely mixed than on the slice \(z=1/2\), and their sampled fractions were approximately \(B_+\simeq0.340\), \(B_-\simeq0.322\), and \(\mathcal L\simeq0.338\). No sampled point was assigned to the residual ``other'' class.
}
\label{fig:rule210_threeclass_z03}
\end{figure}

\FloatBarrier

To quantify the finite-resolution mixing of the three labels, we use an uncertainty-probability calculation inspired by uncertainty-exponent methods for fractal basin boundaries~\cite{GrebogiOttYorke1983}. 
In the present setting, however, the labels are the three finite-resolution classes \(B_+\), \(B_-\), and \(\mathcal L\), rather than two basins alone.
For a fixed slice \(z=z_0\), choose pairs of initial points separated by a distance \(\varepsilon\).
Let
\(\ 
P_{z_0}^{(3)}(\varepsilon)
\ \)
be the probability that the two points receive different labels among
\( \ 
B_+,\ B_-,\ \mathcal L.
\ \)
If, over the fitted finite-resolution range,
\( \ 
P_{z_0}^{(3)}(\varepsilon)\sim\varepsilon^{\gamma_3(z_0)},
\ \)
then
\( \ 
D_{\mathrm{eff}}(z_0)=2-\gamma_3(z_0)
\ \) 
is used as a finite-resolution boundary-complexity estimate. 

\begin{figure}[tb]
\centering
\includegraphics[width=0.95\textwidth]{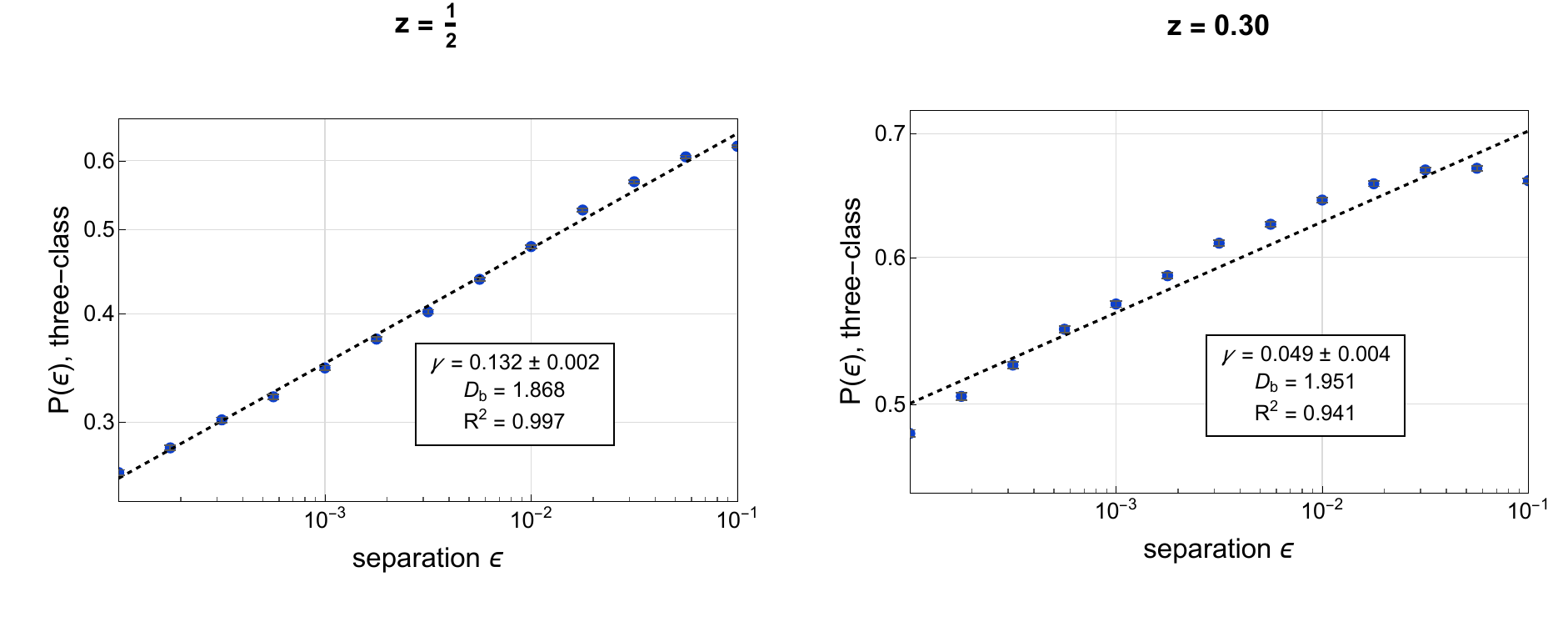}
\caption{
Three-class uncertainty calculation for rule 210 on the slices \(z=1/2\) and \(z=0.3\). 
A pair is counted as uncertain when the two points receive different labels among \(B_+\), \(B_-\), and \(\mathcal L\). 
Error bars indicate binomial standard errors of \(P^{(3)}_{z_0}(\varepsilon)\). 
The value after \(\pm\) is the standard error of the fitted slope. 
The fitted slopes are interpreted as finite-resolution uncertainty slopes over the displayed range, not as rigorous asymptotic exponents. No pair contained a residual ``other'' label under this three-class classification.
}
\label{fig:rule210_threeclass_uncertainty}
\end{figure}

Figure~\ref{fig:rule210_threeclass_uncertainty} shows the three-class uncertainty calculation. Error bars indicate binomial standard errors of \(P_{z_0}^{(3)}(\varepsilon)\). 
The fitted slopes were
\[
\gamma_3(1/2)=0.132\pm0.002,
\qquad
D_{\mathrm{eff}}(1/2)=1.868,
\]
and
\[
\gamma_3(0.3)=0.049\pm0.004,
\qquad
D_{\mathrm{eff}}(0.3)=1.951.
\]
Thus the slice \(z=0.3\) has a smaller finite-resolution uncertainty slope and a larger effective boundary-complexity estimate than the slice \(z=1/2\).

We also compute a finite-resolution three-class entropy. 
This quantity is an analogue of basin entropy~\cite{daza2016Basin}, but it is applied here to the three labels \(B_+\), \(B_-\), and \(\mathcal L\). 
The third label \(\mathcal L\) denotes finite-resolution proximity to the cyclic invariant line set, not a basin of a stable periodic orbit. 
Divide the slice into boxes \(Q_i\), and let \(p_{ir}\) be the fraction of sampled initial conditions in box \(Q_i\) with label
\( \ 
r\in\{+,-,\mathcal L\}.
\ \)
Define
\[
S_i^{(3)}=-\sum_{r\in\{+,-,\mathcal L\}}p_{ir}\log p_{ir},
\quad
\mathrm{and}
\qquad
S_3=\frac1{N_b}\sum_i S_i^{(3)}.
\]
We also compute the entropy restricted to mixed boxes,
\[
S_{bb,3}=\frac1{N_{\mathrm{mix}}}
\sum_{Q_i\ \mathrm{mixed}}S_i^{(3)},
\]
where a box is called mixed if at least two of the three labels occur in it.
The quantities \(S_3\) and \(S_{bb,3}\) are normalized by \(\log 3\).

\begin{figure}[htb]
\centering
\includegraphics[width=0.75\textwidth]{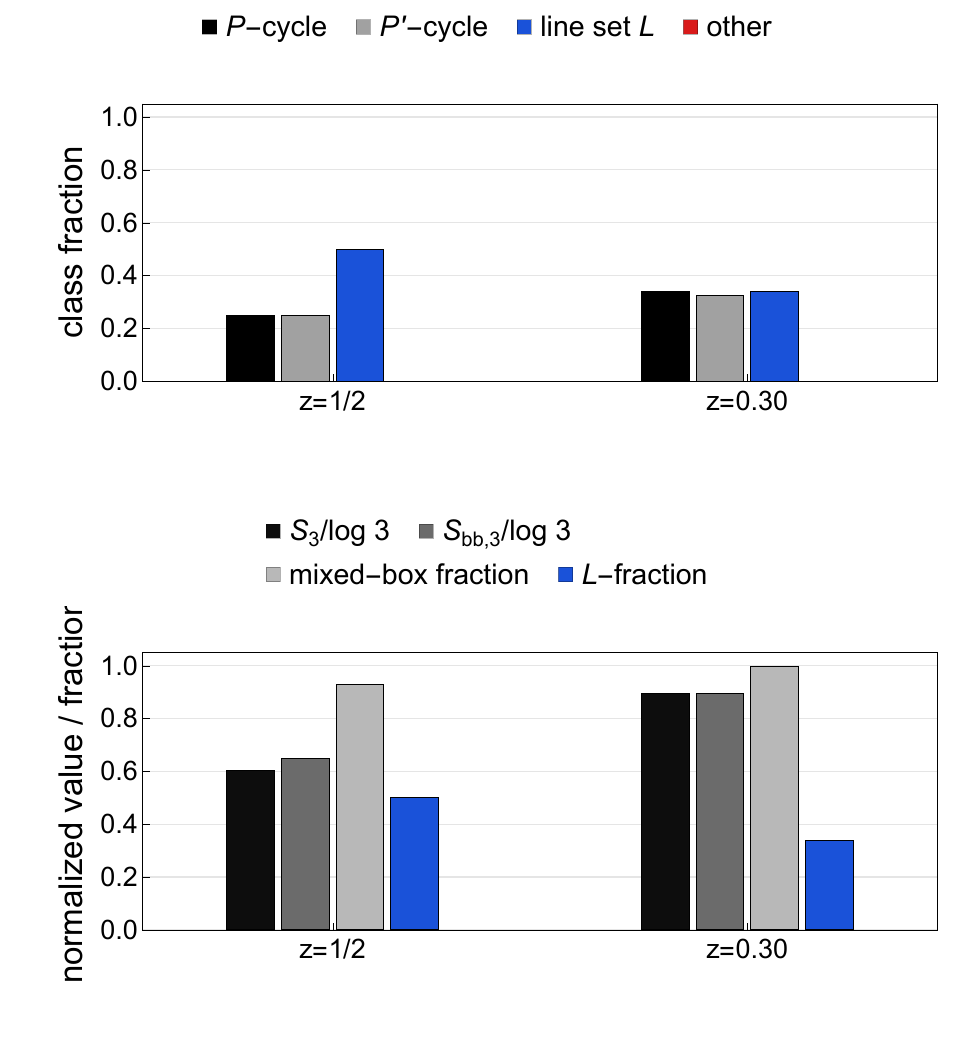}
\caption{
Class fractions and finite-resolution three-class entropy for rule 210.
Upper panel: sampled fractions of the classes \(B_+\), \(B_-\), \(\mathcal L\), and the residual ``other'' label. 
Lower panel: normalized three-class entropy \(S_3/\log 3\), mixed-box entropy \(S_{bb,3}/\log 3\), mixed-box fraction, and \(\mathcal L\)-fraction. 
The main setting uses \(80\times80\) boxes and 20 samples per box. No sampled point was assigned to the residual ``other'' class.
}
\label{fig:rule210_threeclass_summary}
\end{figure}

Figure~\ref{fig:rule210_threeclass_summary} summarizes the class fractions and the three-class entropy. 
In the main setting of \(80\times80\) boxes with 20 samples per box, the normalized entropies were approximately
\[
S_3/\log 3\simeq0.602,
\qquad
S_{bb,3}/\log 3\simeq0.647
\]
on the slice \(z=1/2\), and
\[
S_3/\log 3\simeq0.894,
\qquad
S_{bb,3}/\log 3\simeq0.896
\]
on the slice \(z=0.3\). 
The mixed-box fraction was also larger on the slice \(z=0.3\). 
These results support the visual observation that the three classes are more finely mixed on \(z=0.3\) than on \(z=1/2\).

\FloatBarrier

%%%%%%%%%%%%%%%%%%%%%%%%%%%%%%%%%%%%%%%%%%%%%%%%%%%%%%%%%%%%%%%%%
\subsection{Rules 51 and 85: inherited expanding one-dimensional dynamics}
\label{subsec:rules51_85}

The rule 210 dynamics arises from an interaction between the ECA rule and the expanding fuzzification function. 
In contrast, rules 51 and 85 provide simpler examples in which the three-cell GFECA directly inherits expanding one-dimensional dynamics from \(u_*\).

Under the rule-number convention used here,
\[
f_{51}(x,y,z)=1-y,
\qquad
f_{85}(x,y,z)=1-z.
\]
Define
\(\ 
h_*(x)=1-u_*(x)
\). 
Then the rule 51 three-cell map is
\[
T_{51}(x,y,z)=(h_*(x),h_*(y),h_*(z)).
\]
Thus each coordinate evolves independently under the same one-dimensional piecewise-linear map \(h_*\). On each branch where \(h_*\) is differentiable,
\(\ 
|h_*'(x)|=3\). 
Therefore, away from break points and their preimages, generic orbits have Lyapunov exponent \(\log 3\).
Moreover, \(h_*\) has a period-three orbit; for example,
\[
\frac{1}{26}\mapsto \frac{23}{26}\mapsto \frac{9}{26}
\mapsto \frac{1}{26}.
\]
Thus the inherited one-dimensional dynamics also falls under the period-three-implies-chaos mechanism of Li and Yorke~\cite{LiYorke1975}.

For rule 85, we obtain
\[
T_{85}(x,y,z)=(h_*(y),h_*(z),h_*(x)).
\]
Consequently,
\[
T_{85}^3(x,y,z)=(h_*^3(x),h_*^3(y),h_*^3(z)).
\]
Hence rule 85 inherits the same expanding one-dimensional dynamics after a cyclic permutation of the coordinates.

These two rules illustrate a mechanism different from rule 210. 
In rules 51 and 85, the expanding, Li--Yorke chaotic one-dimensional dynamics is inherited directly from the fuzzification function.
In rule 210, by contrast, the locally stable period-six cycles and the three-class slice structures arise from the interaction between the Boolean ECA rule and the expanding fuzzifier.

%%%%%%%%%%%%%%%%%%%%%%%%%%%%%%%%%%%%%%%%%%%%%%%%%%%%%%%%%%%%%%%%%
\subsection{Summary of three-cell dynamics}
\label{subsec:threecell_summary}

The minimal three-cell setting shows that generalized FDNF can convert a finite Boolean cellular automaton into a genuinely continuous-state nonlinear map. 
The same expanding endpoint-preserving function \(u_*\) produces different mechanisms depending on the underlying ECA rule. 
Rule 210 exhibits two locally stable period-six cycles coexisting with an expanding cyclic invariant line set, and this coexistence produces finely mixed finite-resolution three-class slice structures. 
Rules 51 and 85 inherit expanding one-dimensional dynamics directly from the fuzzification function. 
These examples show that generalized FDNF provides a rule-preserving route from Boolean ECA rules to low-dimensional nonlinear dynamics.

\FloatBarrier

%%%%%%%%%%%%%%%%%%%%%%%%%%%%%%%%%%%%%%%%%%%%%%%%
% sectiion 7
%%%%%%%%%%%%%%%%%%%%%%%%%%%%%%%%%%%%%%%%

\section{Conclusion}

In this paper, we developed a generalized FDNF framework for constructing fuzzy cellular automata from Boolean cellular automata. 
We first interpreted the ordinary FDNF rule as the canonical multi-affine extension of a Boolean local rule to the unit cube. This formulation clarifies both the naturalness of FDNF, which preserves the original Boolean rule on binary states, and its limitation, namely that the resulting fuzzy dynamics is confined to the multi-affine class.

To extend FDNF while retaining its rule-preserving property, we introduced endpoint-preserving transformations. 
For the FDNF polynomial \(f_k\) of an ECA rule \(F_k\), we considered generalized rules of the form
\[
\widetilde f_k^{g,u,v,w}(x,y,z)
=
g\left(f_k(u(x),v(y),w(z))\right),
\]
where \(g,u,v,w:[0,1]\to[0,1]\) preserve the endpoints. Ordinary FDNF is recovered when \(g=u=v=w=\mathrm{id}\), and the generalized rule still agrees with the original ECA rule on \(\{0,1\}^3\). 
Thus, the proposed framework is a structured extension of FDNF based on endpoint-preserving functional composition.

We showed that the choice of these transformations strongly affects pattern formation. 
Threshold-like transformations can recover ECA-like behavior from non-binary initial data, while one-parameter deformations to the identity map describe a transition from ECA-like recovery to FDNF-like smoothing.
Discontinuous transformations with a gap generate additional regimes that are absent in ordinary FDNF. 
These results indicate that ordinary FDNF is one distinguished member of a broader family of rule-preserving fuzzy extensions.

We also introduced quantitative indicators for describing such pattern changes. 
The contrast and fuzziness indices characterize the distribution of cell values, while the effective support exponent, based on normalized high-order moments, measures how broadly non-uniform amplitudes are distributed over a spacetime window. 
Applied to the \(q_a\)-deformation and gap-induced patterns, this exponent provides a compact finite-resolution summary of parameter-dependent amplitude-support changes, with moment-order checks supporting the robustness of the observed trends.

Finally, we examined minimal three-cell GFECA systems. 
Although the corresponding Boolean systems have only finitely many states, their GFECA extensions define continuous maps on the unit cube. Using an expanding non-monotone transformation, we showed that Rule 210 exhibits two explicitly constructible locally stable period-six cycles, verified by interval arithmetic, together with an expanding cyclic invariant line set. 
Rules 51 and 85 illustrate another mechanism, in which expanding one-dimensional dynamics is inherited directly from the fuzzification function. These examples demonstrate that generalized FDNF provides a rule-preserving route from Boolean ECA rules to low-dimensional continuous-state nonlinear dynamics.

Several problems remain open, including a systematic classification of endpoint-preserving transformations, a more complete analysis of the dependence on the underlying ECA rule, and the development of analytical
criteria connecting local transformation properties with global pattern formation. 
For applications, it will also be important to calibrate the choice of transformations to concrete modeling tasks while preserving the structural link to the underlying Boolean cellular automaton.

In summary, the generalized FDNF framework preserves the canonical rule-based structure of FDNF while allowing richer fuzzy and nonlinear dynamics. 
It provides a systematic bridge between Boolean cellular automata, fuzzy cellular automata, and continuous-state nonlinear dynamical systems.

%%%%%%%%%%%%%%%%%%%%%%%%%%%%%%%%%%%%%%%%%%%%%%%%%%%%%
% Appendix
%%%%%%%%%%%%%%%%%%%%%%%%%%%%%%%%%%%%%%%%%%%%%%%%
\appendix

%%%%%%%%%%%%%%%%%%%%%%%%%%%%%%%%%%%%%%%%%%%%%%%%%%%%%%%%%%%%%%%%%
\section{Interval verification of the rule 210 period-six cycle}
\label{app:rule210_interval}

This appendix gives a computer-assisted interval verification of the
period-six cycle of the three-cell rule 210 GFECA used in
Section~\ref{sec:minimal_dynamics}. The accompanying Mathematica script is
included in the supplementary code. The verification uses rational interval endpoints and interval evaluation of
all inequalities below, following the standard framework of interval
arithmetic~\cite{Moore1966}; the software version and run metadata are
recorded in the supplementary data.

We use the rule-number convention defined in Section~2. The FDNF polynomial
of rule 210 is
\[
f_{210}(X,Y,Z)=2XYZ-2XZ+X-YZ+Z.
\]
Let \(u_*\) be defined by \eqref{eq:u_star}, and set
\[
\widetilde f_{210}(x,y,z)=f_{210}(u_*(x),u_*(y),u_*(z)).
\]
The corresponding three-cell map is
\[
T(x,y,z)=
\left(
\widetilde f_{210}(z,x,y),
\widetilde f_{210}(x,y,z),
\widetilde f_{210}(y,z,x)
\right).
\]

We consider the six points
\[
\begin{aligned}
P_0&=(a^*,1/2,b^*),&
P_1&=(c^*,d^*,1/2),&
P_2&=(1/2,b^*,a^*),\\
P_3&=(d^*,1/2,c^*),&
P_4&=(b^*,a^*,1/2),&
P_5&=(1/2,c^*,d^*).
\end{aligned}
\]

The number \(a^*\) is defined as the unique root in
\[
I_a=[0.47853509437,\ 0.47853509439]
\]
of
\[
p_a(t)=94770t^3-211653t^2+149499t-33458.
\]
The signs
\[
p_a(0.47853509437)<0,
\qquad
p_a(0.47853509439)>0
\]
and the interval enclosure
\[
p_a'(I_a)\subset[12038.029627,\ 12038.029631]
\]
prove existence and uniqueness of this root in \(I_a\).

The remaining coordinates are defined by
\[
b^*=
\frac{11664}{281}(a^*)^3
-
\frac{123228}{1405}(a^*)^2
+
\frac{1048793}{18265}a^*
-
\frac{206733}{18265},
\]
\[
c^*=
-\frac{4860}{281}(a^*)^3
+
\frac{10269}{281}(a^*)^2
-
\frac{177843}{7306}a^*
+
\frac{39935}{7306},
\]
and
\[
d^*=
-\frac{12636}{281}(a^*)^3
+
\frac{133497}{1405}(a^*)^2
-
\frac{177843}{2810}a^*
+
\frac{39373}{2810}.
\]
These formulae are obtained from the four algebraic equations
\[
c=9ab-\frac92a-6b+\frac72,
\]
\[
d=-9ab+3a+\frac92b-1,
\]
\[
b=-9cd+6c+\frac32d-\frac12,
\]
and
\[
a=9cd-\frac92c-3d+2.
\]
A Gr\"obner basis calculation gives the three formulae above and the factor
\[
(2a-1)p_a(a)=0.
\]
Since \(I_a\) does not contain \(1/2\), the above definitions solve the four
equations on the relevant branch itinerary.

Interval evaluation gives
\[
a^*\in[0.47853509437,\ 0.47853509439],
\]
\[
b^*\in[0.62355722342,\ 0.62355722347],
\]
\[
c^*\in[0.29079486756,\ 0.29079486759],
\]
and
\[
d^*\in[0.55606665565,\ 0.55606665573].
\]
In particular,
\[
c^*<\frac13<a^*<\frac12<d^*<b^*<\frac23.
\]
Thus the branch assignment is self-consistent:
\[
u_*(c^*)=3c^*,
\]
and
\[
u_*(a^*)=2-3a^*,
\qquad
u_*(b^*)=2-3b^*,
\qquad
u_*(d^*)=2-3d^*.
\]

Substitution on these branches gives
\[
T(P_i)=P_{i+1},\qquad i=0,\ldots,5,
\]
where indices are taken modulo \(6\). Since the intervals for
\(c^*,a^*,1/2,d^*,b^*\) are mutually disjoint, the six points are distinct
and the cycle has minimal period six.

It remains to verify local stability. Let
\[
M=DT(P_5)DT(P_4)\cdots DT(P_0)
\]
be the Jacobian matrix of \(T^6\) at \(P_0\). Interval evaluation of the
characteristic polynomial
\[
\lambda^3+A\lambda^2+B\lambda+C
\]
gives
\[
A\in[-0.747555425,\ -0.747555334],
\]
\[
B\in[0.455938180,\ 0.455938391],
\]
and
\[
C\in[-0.001289829,\ -0.001289826].
\]
The Jury quantities are enclosed by
\[
1+A+B+C\in[0.707092927,\ 0.707093230],
\]
\[
1-A+B-C\in[2.204783341,\ 2.204783644],
\]
and
\[
1-B+AC-C^2\in[0.545024162,\ 0.545024375].
\]
Moreover,
\[
|C|<0.001290<1.
\]
By the Jury stability criterion~\cite{Jury1964} for a monic cubic
polynomial, all roots of the characteristic polynomial lie strictly inside
the unit disk.
Hence the period-six
orbit \(P_0,\ldots,P_5\) is locally asymptotically stable. The symmetric
period-six orbit \(P'_i=\rho^i\kappa(P_i)\) has the same stability by the
complement-reversal symmetry.

%%%%%%%%%%%%%%%%%%%%%%%%%%%%%%%%%%%%%%%%%%%%%%%%%%%%%%%%%%%%%%%%%
\section{Numerical protocols and supplementary checks}
\label{app:numerical_protocols}

This appendix summarizes the numerical protocols used for the quantitative
figures. The Mathematica scripts, summary CSV files, initial-condition files,
and metadata files are included in the supplementary data.

\subsection{Contrast and fuzziness for Figs.~8 and 10}
\label{app:RB_protocols}

For the \(q_a\)-deformation contrast/fuzziness experiment in Fig.~8, the
experiments were performed for rules \(30,90,184\) on a periodic lattice of
\(N=100\) sites up to \(T=100\). The parameter grid was
\[
a=0,0.01,\ldots,1.00.
\]
For each rule, the same ensemble of 50 i.i.d. uniform random initial
conditions in \([0,1]^N\) was used for all parameter values. The random seed
was 20260316. The plotted curves are ensemble means of
\[
R_k(a;T)=\max_j x_j^T-\min_j x_j^T
\]
and
\[
B_k(a;T)=\frac1N\sum_{j=0}^{N-1}4x_j^T(1-x_j^T).
\]
Gray bands indicate one standard deviation.

For the gap-function contrast/fuzziness experiment in Fig.~10, the experiments
were performed for rules \(102\) and \(184\) on a periodic lattice of \(N=50\)
sites up to \(T=200\). The parameter grid was
\[
a=1.00,1.01,\ldots,2.00.
\]
The metrics were computed on the late-time window
\[
D=\{(t,j)\mid100\le t\le200,\ 0\le j<N\}.
\]
For each rule, the same ensemble of 30 i.i.d. uniform random initial
conditions in \([0,1]^N\) was used for all parameter values. The random seed
was 20260316. The plotted curves are ensemble means of
\[
R_k(a)=\max_{(t,j)\in D}x_j^t-
\min_{(t,j)\in D}x_j^t
\]
and
\[
B_k(a)=\frac1{|D|}\sum_{(t,j)\in D}4x_j^t(1-x_j^t).
\]

\subsection{Moment-order robustness of the support exponent}
\label{app:moment_order_robustness}

To check the dependence of the support exponent on the moment order, we
recomputed \(s^{(p)}\) for
\[
p=8,16,32,64
\]
using the same spacetime windows, initial conditions, and rounding convention
as in Table~\ref{tab:support_settings}. We also recorded the amplitude
normalization
\[
S_2/M
\]
for each parameter value. The qualitative parameter dependence of the support
exponent was unchanged under these choices of \(p\), although the absolute
values varied as expected. Parameter ranges in which \(S_2/M\) is extremely
small should be interpreted with caution, because the support exponent is
invariant under multiplication of all amplitudes by a common constant.

\subsection{Three-class classification protocol for rule 210}
\label{app:rule210_threeclass_protocol}

For the rule 210 slice classifications in
Figs.~\ref{fig:rule210_threeclass_z05}--\ref{fig:rule210_threeclass_summary},
we used the three labels \(B_+\), \(B_-\), and \(\mathcal L\), together with
a residual diagnostic label ``other''. The convergence criterion to a period-six
cycle used tolerance
\[
\delta=10^{-6},
\]
maximum iteration time
\[
n_{\max}=20000,
\]
and required 20 consecutive checks at period-six spacing. The line-set proximity
test used the distance \(d_{\mathcal L}\) in \eqref{eq:line_set_distance}. The
slice images were computed on an \(800\times800\) grid. No sampled point was
assigned to the residual ``other'' class.

For the uncertainty calculation, pairs of points at separation \(\varepsilon\)
were sampled in the slice. A pair was counted as uncertain when the two
points received different labels among
\[
B_+,\quad B_-,\quad \mathcal L.
\]
Error bars in Fig.~\ref{fig:rule210_threeclass_uncertainty} are binomial
standard errors. The fitted slopes were obtained by least-squares fitting in
log-log coordinates over the displayed \(\varepsilon\)-range.

For the class entropy calculation, the main setting used \(80\times80\) boxes
and 20 samples per box. Resolution checks using different numbers of boxes
and samples per box are provided in the supplementary data.

\subsection{Resolution checks for the three-class entropy}
\label{app:rule210_resolution_checks}

\begin{figure}[tb]
\centering
\includegraphics[width=0.82\textwidth]{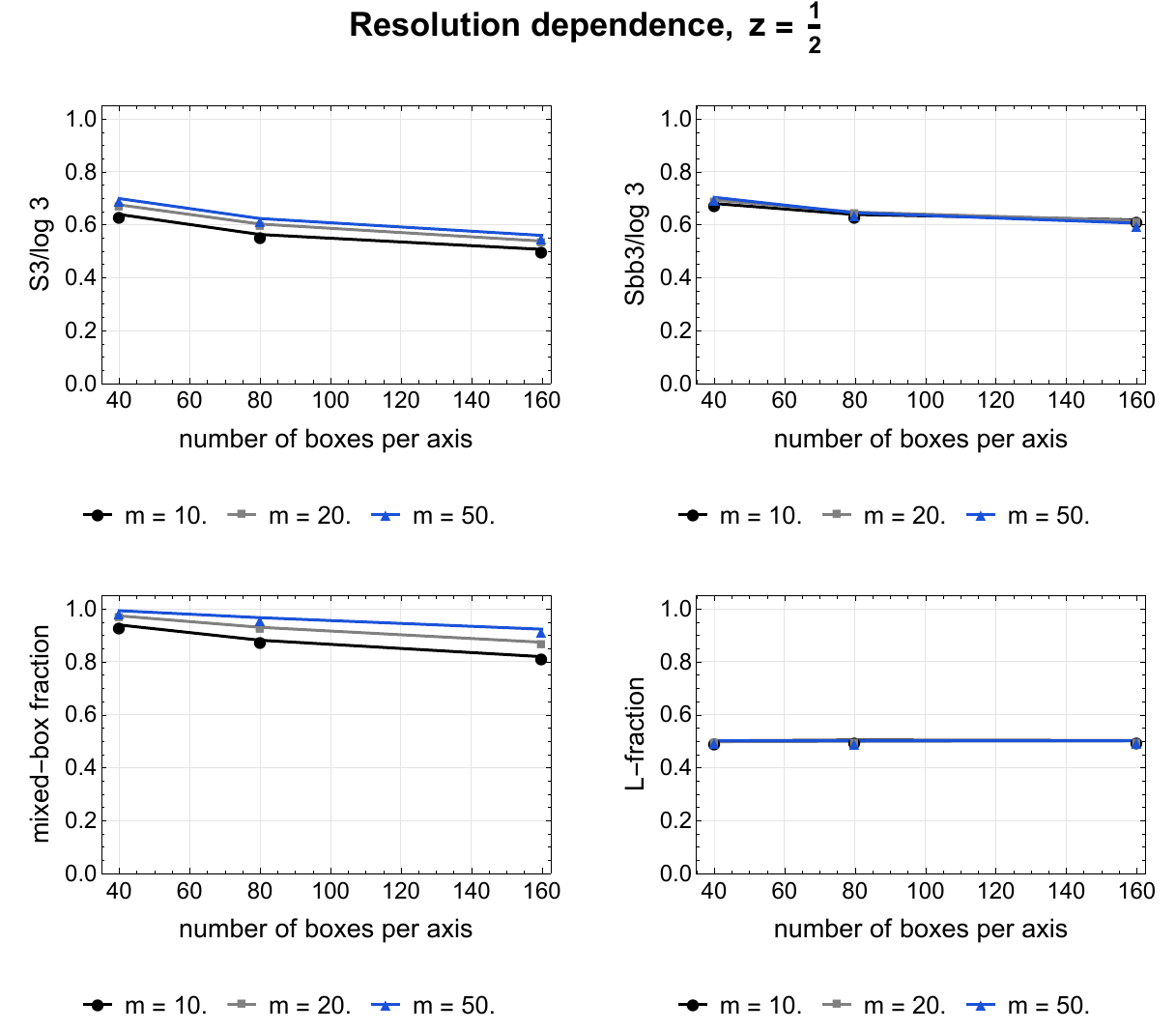}
\caption{
Resolution check of the three-class entropy and \(\mathcal L\)-fraction for
rule 210 on the slice \(z=1/2\). The horizontal axis is the number of boxes
per coordinate direction. Different curves correspond to different numbers of
sampled points per box. These plots are provided as a finite-resolution
robustness check for Fig.~\ref{fig:rule210_threeclass_summary}.
}
\label{fig:rule210_resolution_z05}
\end{figure}

\begin{figure}[tb]
\centering
\includegraphics[width=0.82\textwidth]{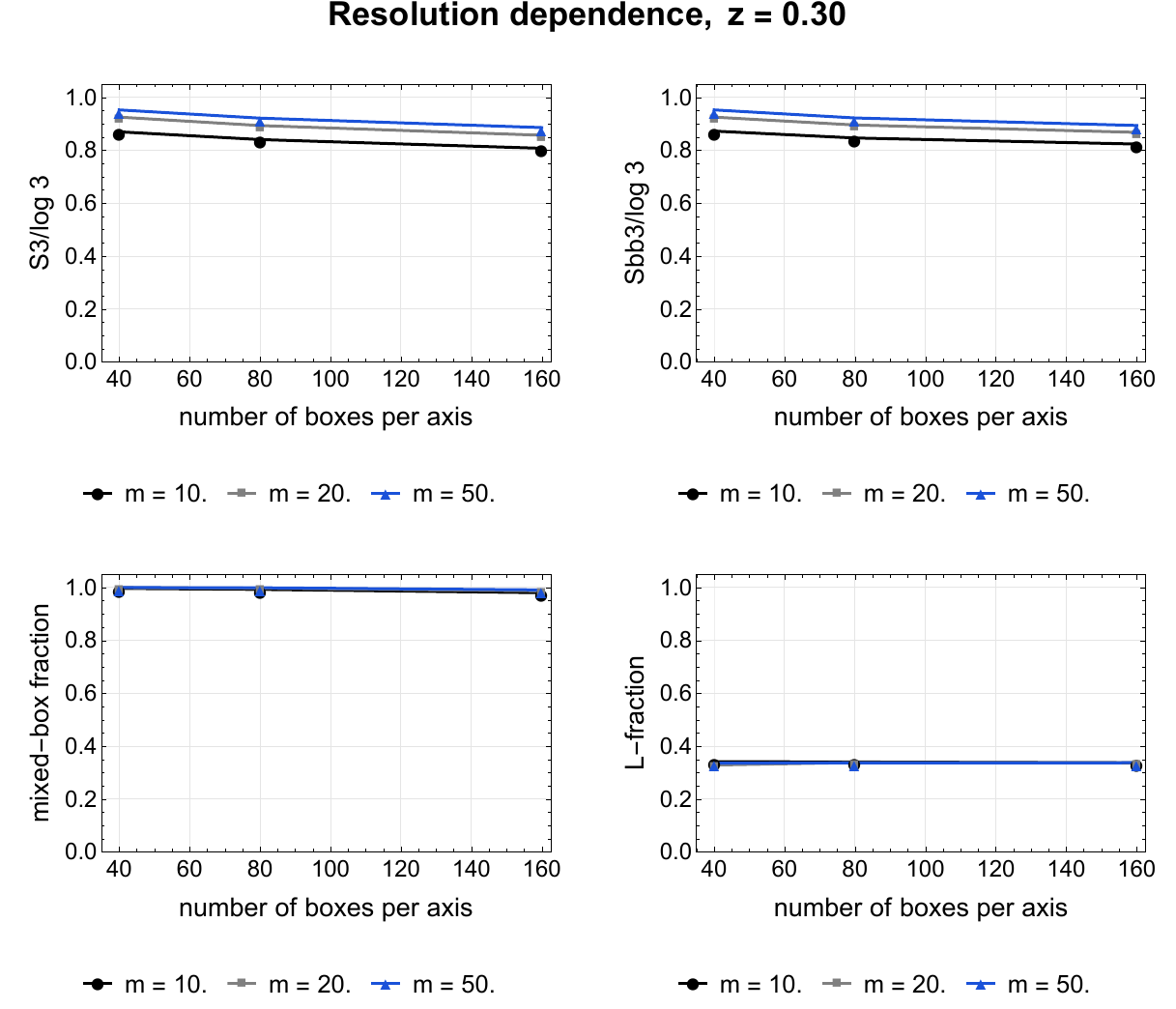}
\caption{
Resolution check of the three-class entropy and \(\mathcal L\)-fraction for
rule 210 on the slice \(z=0.3\). The same protocol as in
Fig.~\ref{fig:rule210_resolution_z05} is used. The values remain consistently
larger than those on the slice \(z=1/2\), indicating stronger finite-resolution
mixing of the three classes.
}
\label{fig:rule210_resolution_z03}
\end{figure}

The main class-entropy calculation in Fig.~\ref{fig:rule210_threeclass_summary}
uses \(80\times80\) boxes and 20 samples per box. To check that the qualitative
ordering of the two slices is not an artifact of this particular resolution, we
also computed the same quantities for several box resolutions and numbers of
samples per box. Figures~\ref{fig:rule210_resolution_z05} and
\ref{fig:rule210_resolution_z03} show the resulting resolution checks for the
slices \(z=1/2\) and \(z=0.3\), respectively. In all tested settings, the slice
\(z=0.3\) has larger three-class entropy and mixed-box fraction than the slice
\(z=1/2\), consistent with the main-text classification plots.

\FloatBarrier

%%%%%%%%%%%%%%%%%%%%%%%%%%%%%%%%%%%%%%%%%%%%%%%%%%%%%%%%%%%%%%%%%
\section{Illustrative convex mixing beyond the single-rule GFDNF setting}
\label{app:convex_mixing}

This appendix presents an illustrative extension in which two generalized fuzzy rules are combined by a convex mixture. This construction is not, in general, an endpoint-preserving fuzzification of a single Boolean ECA rule. Therefore, it is not used as a main component of the single-rule GFDNF framework developed in the main text. Instead, it is included only to show that generalized fuzzy rules can also be combined to produce branch-dependent changes in pattern formation.

We consider
\[
\widetilde f_{184,90}^{(\alpha)}(x,y,z)
=
(1-\alpha)\widetilde f_{184}(x,y,z)
+
\alpha \widetilde f_{90}(x,y,z),
\qquad
0\le\alpha\le1.
\]
Here \(\widetilde f_{184}\) and \(\widetilde f_{90}\) are the generalized FDNF rules constructed with
\[
g=\mathrm{id},
\qquad
u=v=w=q_0.
\]
Thus \(\alpha=0\) corresponds to the rule 184 fuzzy rule and \(\alpha=1\) corresponds to the rule 90 fuzzy rule.

For each \(\alpha\), we compute the spacetime pattern from the localized initial condition
\[
x_0^0=0.6, \qquad x_j^0=0 \quad (j\ne0),
\]
on a periodic lattice with \(N=50\) up to \(T=500\). The parameter grid is \(0\le\alpha\le1\) with \(\Delta\alpha=0.005\). The support exponent is evaluated on the late-time window
\[
D=\{(t,j)\mid 450\le t\le500,\ 0\le j<N\}
\]
using the field \(\phi_{t,j}=x_j^t\) and moment order \(p^*=64\). This is a single-run illustrative experiment, and the values should not be directly compared with the \(p^*=16\) ensemble results in the main text.

\begin{figure}[htb]
\centering
\includegraphics[width=0.75\textwidth]{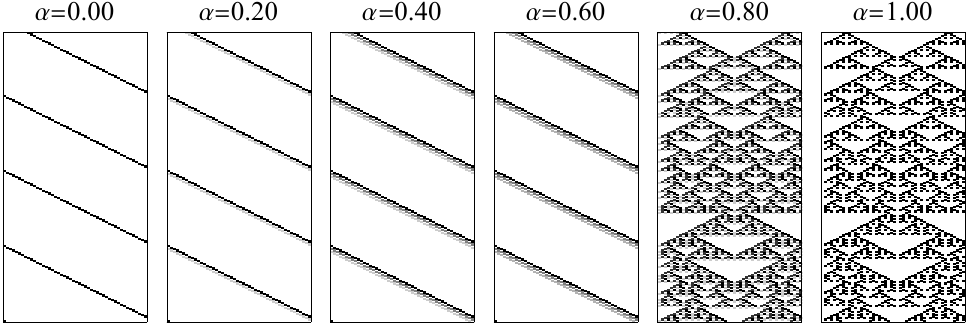}
\caption{
Time-evolution patterns generated by the convex mixture of rule 184 and rule 90. The displayed time interval is \(0\le t\le 200\), the system size is \(N=50\), and the initial condition is \(x_0^0=0.6\) with all other cells set to zero. The columns correspond to \(\alpha=0,0.2,0.4,0.6,0.8,1.0\). The support exponent in Fig.~\ref{fig:mixing_support} is computed on the later window specified above.
}
\label{fig:mixing_patterns_overview}
\end{figure}

\begin{figure}[htb]
\centering
\includegraphics[width=0.75\textwidth]{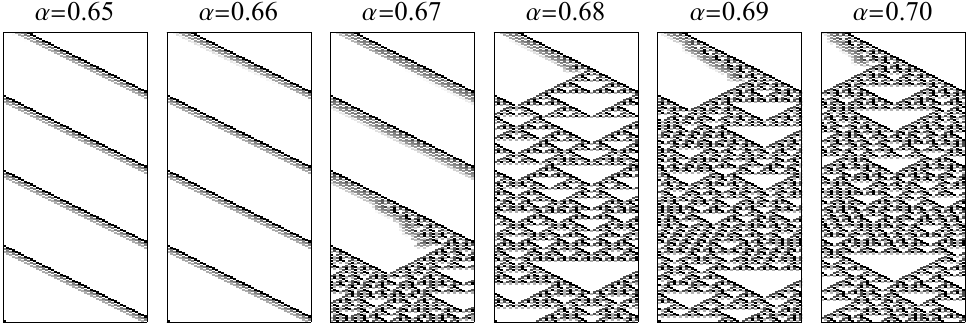}
\caption{Magnified view of the mixing dynamics near the branch-switching crossover. The columns correspond to \(\alpha=0.65,0.66,0.67,0.68,0.69,0.70\).}
\label{fig:mixing_patterns_zoom}
\end{figure}

\begin{figure}[htb]
\centering
\includegraphics[width=0.5\textwidth]{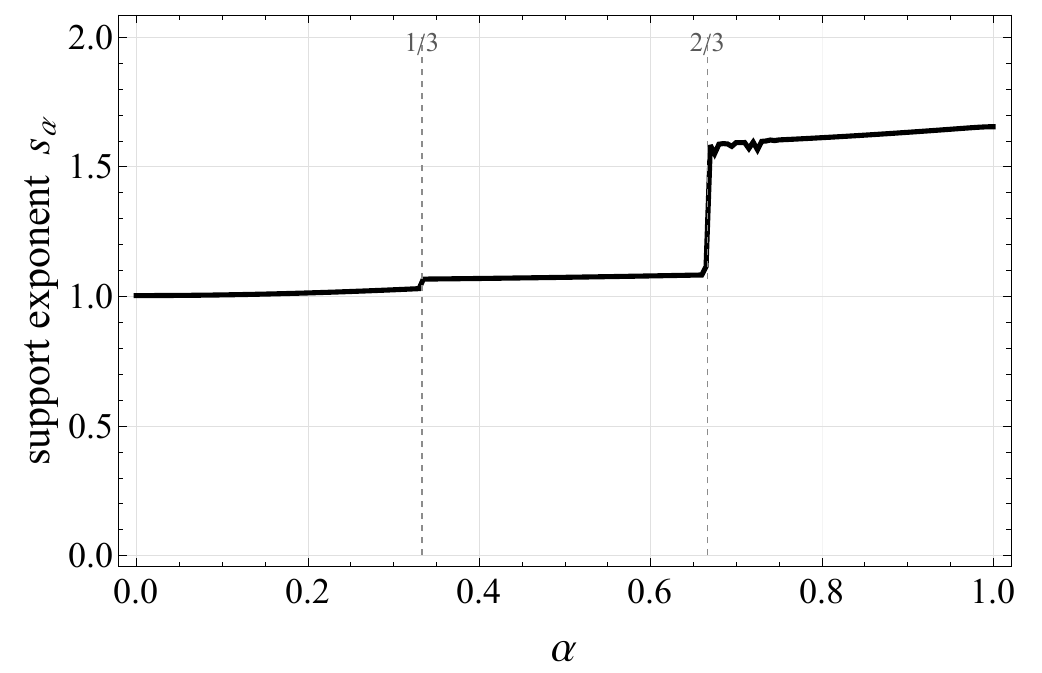}
\caption{Effective support exponent for the convex mixture of rule 184 and rule 90 fuzzy rules. This single-run illustrative experiment uses \(p^*=64\), so the values should not be directly compared with the ensemble results computed with \(p^*=16\). The parameter \(\alpha\) interpolates from rule 184 at \(\alpha=0\) to rule 90 at \(\alpha=1\). Dashed vertical lines indicate \(\alpha=1/3\) and \(\alpha=2/3\).}
\label{fig:mixing_support}
\end{figure}

The curve displays a sharp parameter-dependent crossover. We refer to this as a branch-switching crossover rather than a thermodynamic phase transition. The branch-switching interpretation is motivated by the threshold-like transformation \(q_0\), whose branches meet at \(1/3\) and \(2/3\). Since \(q_0\) maps values below \(1/3\) to \(0\) and values above \(2/3\) to \(1\), the mixed outputs \(\alpha\) and \(1-\alpha\) can change their symbolic images when \(\alpha\) crosses \(1/3\) or \(2/3\). As \(\alpha\) varies, the values generated by the convex mixture may cross these branch values, producing a rapid change in the subsequent symbolic pattern.

This mixing experiment is intended to illustrate how generalized fuzzy rules can be combined to produce branch-dependent changes in pattern formation. A more systematic finite-size and initial-condition study would be needed to assess the robustness of this crossover.

\FloatBarrier

\section*{CRediT authorship contribution statement}
All authors contributed equally to the study conception
and design. The first draft of the manuscript was written by T.T., and all authors
commented on previous versions of the manuscript. All authors read and approved
the final manuscript.
\section*{Funding}
This work was supported by JSPS KAKENHI Grant Number
JP23K22408, JP24K06819 and JP26K06932. It was also supported by Arithmer Inc.
%%%
\section*{Data availability}
The numerical data and code supporting the figures and reproducibility
checks are provided as Supplementary Data accompanying this article.
\section*{Declaration of generative AI and AI-assisted technologies in the manuscript preparation process}
During the preparation of this work, the authors used ChatGPT to assist
with language editing, manuscript organization, and file-structure checking.
After using this tool, the authors reviewed and edited the content as needed
and take full responsibility for the content of the published article.

\section*{Acknowledgements}
The authors would like to thank Prof. Junkichi Satsuma and Dr. Kohei Higashi for useful discussions.

  \bibliographystyle{elsarticle-num} 
  \bibliography{ref_shimizu}



%\begin{thebibliography}{00}

%% For numbered reference style

%\end{thebibliography}
\end{document}